\documentclass[epj,nopacs]{svjour}
\usepackage{graphicx,amsmath,amssymb,bm}
\usepackage[applemac]{inputenc}
\usepackage{epsfig}
\usepackage{longtable}
\usepackage{nicefrac}
\usepackage{cite}

\usepackage{color}
\definecolor{purple}{rgb}{0.5,0,0.5}
\definecolor{blue}{rgb}{0.0,0,0.9}
\usepackage[colorlinks=true, pdfstartview=FitV, linkcolor=purple, citecolor= purple,urlcolor=blue]{hyperref}
\usepackage{tabularx}

\newcolumntype{L}[1]{>{\raggedright\arraybackslash}p{#1}}
\newcolumntype{C}[1]{>{\centering\arraybackslash}p{#1}}
\newcolumntype{R}[1]{>{\raggedleft\arraybackslash}p{#1}}

\begin{document}

\title{Distribution Amplitudes of Heavy Mesons and Quarkonia on the Light Front}

\author{
Fernando~E.~Serna\inst{1,2} \and Roberto~Correa~da~Silveira\inst{1} \and J.~J.~Cobos-Martínez\inst{3,4} \and Bruno~El-Bennich\inst{1,2} \and Eduardo~Rojas\inst{5} }
%%  \thanks{\emph{Present address:} Insert the address here if needed}%
\institute{ LFTC, Universidade Cidade de São Paulo, Rua Galvão Bueno, 868, 01506-000 São Paulo, São Paulo, Brazil
\and 
IFT, Universidade Estadual Paulista, Rua Dr.~Bento Teobaldo Ferraz, 271, 01140-070 S\~{a}o Paulo, SP, Brazil. 
\and
Departamento de F\'isica, Universidad de Sonora, Boulevard Luis Encinas J. y Rosales, Hermosillo, Sonora 83000, M\'exico
\and
Cátedra CONACyT, Departamento de Física, Centro de Investigación y de Estudios Avanzados del Instituto Politécnico Nacional, Apartado Postal 14-740, 
7000, Ciudad de México, México.
\and
Departamento de Física, Universidad de Nariño, A.A. 1175, San Juan de Pasto, Colombia  }

\date{Received: \today}% / Revised version: }

\abstract{The ladder kernel of the Bethe-Salpeter equation is amended by introducing a different flavor dependence of the dressing functions
in the heavy-quark sector. Compared with earlier work this allows for the simultaneous calculation of the mass spectrum and leptonic decay 
constants of light pseudoscalar mesons, the $D_u$, $D_s$, $B_u$, $B_s$ and $B_c$ mesons and the heavy quarkonia $\eta_c$ and $\eta_b$ 
within the same framework at a physical pion mass. The corresponding Bethe-Salpeter amplitudes are projected onto the light front and we reconstruct 
the distribution amplitudes of the mesons in the full theory. A comparison with the first inverse moment of the heavy meson distribution amplitude 
in heavy quark effective theory is made.
  %end of keyword  
\vspace*{-2mm}
 % end of PACS codes
}  %end of abstract

\maketitle

%%%%%%%%%%%%%%%%%%%%%%%%%%%%%%%%%%%%%%%%%%%%%%%%%%%%%%%%%%%%%%%%%%%%%%%%%%%%%%%%%%%%%%%%%%
%%%%%%%%%%%%%%%%%%%%%%%%%%%%%%%%%%%%%%%%%%%%%%%%%%%%%%%%%%%%%%%%%%%%%%%%%%%%%%%%%%%%%%%%%%

\section{Introduction}
\label{intro}

The introduction of hadronic light-cone distribution amplitudes (LCDA) dates back to the seminal works on hard exclusive reactions in perturbative 
QCD~\cite{Chernyak:1977fk,Efremov:1978rn,Efremov:1979qk,Lepage:1979zb,Lepage:1980fj}. These nonperturbative and scale-dependent functions can be understood as 
the closest relative of quantum mechanical wave functions in quantum field theory. They describe the longitudinal momentum distribution of valence quarks in a hadron in the 
limit of negligible transverse momentum, here given by the leading Fock-state contribution to its light-front wave function, the so-called leading-twist LCDA. In particular, 
the light-front formulation of a wave function allows for a probability interpretation of partons not readily accessible in an infinite-body field theory, since particle number 
is conserved in this frame. In other words,  $\phi(x,\mu)$ expresses the light-front fraction of the hadron's momentum carried by a valence quark. 

The simplest hadronic distribution amplitude is that of the pion and has justifiably received much attention~\cite{Ball:2005ei,Braun:2006dg,Arthur:2010xf,Chang:2013pq,
Stefanis:2014nla,Stefanis:2015qha,Shi:2015esa,deMelo:2015yxk,deMelo:2016uwj,Radyushkin:2017gjd,Braun:2015axa,Bali:2017gfr,Bali:2018spj,Bali:2019dqc} 
due to increasing interest in, amongst others, precision calculation of two-photon transition form factor~\cite{Radyushkin:1996tb,Braun:2007wv,Brodsky:2011yv,Masjuan:2012wy,
ElBennich:2012ij,deMelo:2013zza,Raya:2015gva,Mikhailov:2016klg,Choi:2017zxn,Raya:2019dnh} and of weak semi-leptonic, $B \to \pi \ell\nu_\ell$, and non-leptonic 
$B\to \pi\pi, B\to \rho\pi, B\to K\pi ...$ decays. The latter can also be treated as a hard exclusive process with associated factorization theorem, which separates the 
decay amplitudes for a given process into hard short-distance contributions and soft nonperturbative matrix elements, where the distribution amplitudes enter both, 
the hard-scattering integrals over Wilson coefficients and the heavy-to-light transition amplitudes~\cite{Beneke:1999br,Beneke:2000ry,Beneke:2002jn,
Bauer:2000yr,Bauer:2001yt,Bauer:2005wb,ElBennich:2006yi,ElBennich:2009da,Leitner:2010fq}. 

The slowly establishing consensus, after decade-long controversies about the shape of pion's distribution amplitude, points at a function $\phi_\pi(x,\mu)$ that is a concave 
function symmetric about $x=1/2$ and broader than the asymptotic distribution $\phi(x,\mu) \xrightarrow{ \mu\to \infty } 6x(1-x)$~\cite{Bali:2019dqc,Cui:2020dlm}, 
where $x$ is the  longitudinal light-front momentum fraction and $\mu$ is the renormalization scale. For heavier quarkonia, such as the $\eta_c$ and $\eta_b$, the distribution 
amplitudes appear to be increasingly more localized in $x\in[0,1]$ with narrower width and a convex-concave functional behavior~\cite{Bondar:2004sv,Ebert:2006xq,Ding:2015rkn}. 
In the infinite-mass limit these distribution amplitudes tend towards a $\delta$-like function, though this limit is far from being reached at the bottom-mass scale. 
The transition from concavely shaped distribution amplitudes to convex-concave ones occurs between the strange and charm quark, a mass-scale region
known for the onset of important flavor-symmetry breaking effects~\cite{ElBennich:2010ha,ElBennich:2011py,ElBennich:2012tp,El-Bennich:2016bno,El-Bennich:2017brb}.

With respect to factorization approaches in weak heavy-meson decays, distribution amplitudes of heavy mesons defined in heavy quark effective theory~\cite{Grozin:1996pq}
were for the longest time based on models whose functional form in a given limit is dictated by QCD sum rules~\cite{Grozin:1996pq,Braun:2003wx,Ball:2003fq,Khodjamirian:2005ea}, 
guided by an operator product expansion~\cite{Lee:2005gza} or obtained from a combination of dispersion relations and light-cone QCD sum rules~\cite{Beneke:2018wjp}.
Additional model approaches exist, see Refs.~\cite{Bell:2008er,Bell:2013tfa,Wu:2013lga,Tang:2019gvn} for instance. A heavy-light LCDA was also extracted from the extrapolation
of Bethe-Salpeter amplitudes calculated with an unphysical pion mass~\cite{Binosi:2018rht}. 

Herein we re-appreciate earlier work on heavy-light mesons and quarkonia~\cite{Rojas:2014aka,El-Bennich:2016qmb,Mojica:2017tvh,Bedolla:2015mpa,Raya:2017ggu,
Fischer:2014cfa,Hilger:2017jti} within a continuum approach to two-point and four-point functions whose salient features will be summarized in Section~\ref{DSE-BSE}. 
The crucial difference in the present approach is the  flavor-dependence of the interaction in the ladder truncation of the Bethe-Salpeter kernel, 
as we effectively take into account that the  quark-gluon vertex dressing has a different impact for a light quark than for a charm or bottom quark. In general, $D$ and 
$B$ mesons are of particular interest as they offer a rich laboratory to study two limiting mass-scale sectors of QCD with associated emergent approximate symmetries: 
chiral symmetry in the sector of light quarks where  $m_q \ll \Lambda_\mathrm{QCD}$ and heavy quark symmetry for masses $m_q \gg \Lambda_\mathrm{QCD}$
\cite{ElBennich:2012tp,Manohar:2000dt}.  In Ref.~\cite{Serna:2017nlr} we applied these considerations to a contact-interaction model to obtain the mass spectrum 
and decay constants of $D$ mesons. It turned out that introducing different effective couplings, due to unlike dressing effects for  light and heavy 
quarks, in the ladder truncation of the Bethe-Salpeter kernel significantly improved the description of experimental results. This was also observed in 
Ref.~\cite{Gutierrez-Guerrero:2019uwa}.

This logic was subsequently applied to the Bethe-Salp\-eter equation (BSE)~\cite{Chen:2019otg,Qin:2019oar} with a flavor-dependent infrared component of the interaction 
model introduced in Ref.~\cite{Qin:2011dd}. We employ the combined approach of the Dyson-Schwinger equation (DSE) for the quark and BSE with a flavor-dependent, slightly modified 
interaction to first compute the mass spectrum and weak decay constants of the pseudoscalar $\pi$, $K$, $D$, $D_s$, $B$, $B_s$ and $B_c$ mesons 
and $\eta_c$ and $\eta_b$ quarkonia in Section~\ref{DSE-BSE}. In doing so, we fix the light and heavy quark flavors uniquely and solve the DSE on the complex plane 
using Cauchy's theorem~\cite{Fischer:2005en,Krassnigg:2009gd}. The resulting nonperturbative propagators are then inserted consistently and simultaneously in the BSE
for the aforementioned mesons. Since the resulting masses and decay constants are obtained without any extrapolations of eigenvalues or masses, 
we also obtain the corresponding LCDA at a physical mass with appropriate projections of the Bethe-Salpeter amplitudes on the light front in Sections~\ref{pda} 
and \ref{results}. These distributions amplitudes are then used to compute their first inverse moment which we compare with values in heavy quark effective theory in 
Section~\ref{HQET}. We wrap up with a brief conclusion in Section~\ref{remarks}.

%%%%%%%%%%%%%%%%%%%%%%%%%%%%%%%%%%%%%%%%%%%%%%%%%%%%%%%%%%%%%%%%%%%%%%%%%%%%%%%%%%%%%%%%%%
%%%%%%%%%%%%%%%%%%%%%%%%%%%%%%%%%%%%%%%%%%%%%%%%%%%%%%%%%%%%%%%%%%%%%%%%%%%%%%%%%%%%%%%%%%

\section{Pseudoscalar Bound States \label{DSE-BSE}}

The calculation of a meson's distribution amplitude requires the knowledge of the wave function of this bound state. We do so regarding the mesons 
as a continuum bound-state problem described by the homogeneous BSE in leading symmetry-preserving truncation. The solution of this eigenvalue problem yields the mass 
and the Bethe-Salpeter amplitude (BSA) of the meson which can be projected on the light-front to extract a distribution amplitude. It also allows to obtain the leptonic decay 
constant which directly tests the wave function normalization of the meson. The main ingredients of the BSE's kernel are the dressed-quark propagators and the dressed 
effective gluon interaction which are described in the next sections.

%%%%%%%%%%%%%%%%%%%%%%%%%%%%%%%%%%%%%%%%%%%%%%%%%%%%%%%%%%%%%%%%%%%%%%%%%%%%%%%%%%%%%%%%%%

\subsection{Dressed-Quark Propagators}

The dressed propagators are solutions of the quark's gap equation which can be obtained from the appropriate DSE for a given flavor. The DSE describes the 
two-point Green function in terms of a non-linear tower of coupled integral equations, each involving other Green functions, most prominent amongst them the dressed-gluon 
propagator and the quark-gluon vertex. The DSE for a quark of flavor $f$ reads~\cite{Bashir:2012fs,Cloet:2013jya}, \hspace*{-2mm}
%%%%%%%%%%%%%%%%%%%%%%%%%%%%%%%%%%%%%%%%%%%%%%%%%%%%%%%%%%%%%%%%%%%%%%%%%%%%%%%%%%%
\footnote{Henceforth we employ a Euclidean metric which implies:  $\{\gamma_\mu,\gamma_\nu\} = 2\delta_{\mu\nu}$; $\gamma_\mu^\dagger = \gamma_\mu$;  
$\gamma_5= \gamma_4\gamma_1\gamma_2\gamma_3$, tr$[\gamma_4\gamma_\mu\gamma_\nu\gamma_\rho\gamma_\sigma]=-4\, \epsilon_{\mu\nu\rho\sigma}$; 
$\sigma_{\mu\nu}=(i/2)[\gamma_\mu,\gamma_\nu]$;  $a\cdot b = \sum_{i=1}^4 a_i b_i$; and for a time-like vector $P_\mu$ we have $\Rightarrow$ $P^2<0$. } 
%%%%%%%%%%%%%%%%%%%%%%%%%%%%%%%%%%%%%%%%%%%%%%%%%%%%%%%%%%%%%%%%%%%%%%%%%%%%%%%%%%%
%
\begin{align}
\hspace*{-1.5mm}
S^{-1}_f (p)  & =  \, Z_2^f  \left (i\, \gamma \cdot  p + m^{\mathrm{bm}}_f \right )  \nonumber \\
                    & +  \, Z_1^f g^2 \!\! \int^\Lambda  \hspace*{-1.5mm} \frac{d^4k}{(2\pi)^4}  \, D^{ab}_{\mu\nu} (q) \frac{\lambda^a}{2} \gamma_\mu S_f(k) \Gamma^b_{\nu,f}  (k,p) \, ,
\label{QuarkDSE}
\end{align}
where $m^\textrm{bm}_f$ is the bare current-quark mass, $Z_1^f(\mu,\Lambda)$ and $Z_2^f(\mu,\Lambda)$ are the vertex and  wave-function renormalization 
constants at the renormalization point $\mu$, respectively. The integral is over the dressed-quark propagator $S_f(k)$, the dressed-gluon propagator $D_{\mu\nu}(q)$ 
with momentum $q=k-p$ and the quark-gluon vertex, $\Gamma^a_\mu (k,p) = \frac{1}{2}\,\lambda^a \Gamma_\mu (k,p) $, where  the color SU(3) matrices $\lambda^a$ 
are in the fundamental  representation; $\Lambda$ is a Poincaré-invariant regularization scale, chosen such that $\Lambda \gg \mu$. 

The most general Poincaré covariant form of the solutions to Eq.~\eqref{QuarkDSE} is written in terms of scalar and vector contributions:  
\begin{align}
\label{DEsol}
   S_f (p) & =  \, -i \gamma \cdot p \, \sigma_{\rm v}^f ( p^2 ) + \sigma_{\rm s}^f ( p^2 ) \nonumber \\
      & =   \, 1 / \left [ i \gamma \cdot p \,A_f (p^2)   + B_f ( p^2 ) \right  ] \nonumber  \\
                       & =   \, Z_f (p^2 ) / \left [ i \gamma \cdot p + M_f ( p^2 ) \right  ] \ .
\end{align}
The scalar and vector dressing functions are $\sigma_{\rm s}^f ( p^2 )$ and $\sigma_{\rm v}^f ( p^2 )$, respectively, whereas $Z_f(p^2)$ defines the quark's wave function and 
$M_f (p^2) = B_f (p^2)/A_f (p^2)$ is the running mass function. In a subtractive renormalization scheme the two renormalization conditions,
\begin{align}
   Z_f (p^2) & =   \left. 1/A_f (p^2)  \right |_{p^2 = \mu^2} = 1 \  ,
\label{EQ:Amu_ren}   \\
   S^{-1}_f (p)  &     \left.  \right |_{p^2=\mu^2}   =  \,  i \gamma\cdot p \ + m_f(\mu ) \ ,
 \label{massmu_ren}
\end{align}
are imposed, where $m_f(\mu )$ is the renormalized current-quark mass related to the bare mass by,
\begin{equation}
\label{mzeta} 
   Z_4^f  (\mu,\Lambda )\, m_f(\mu)  = Z_2^f  (\mu,\Lambda ) \, m_f^{\rm bm} (\Lambda) \  ,
\end{equation}
 and $Z_4^f(\mu,\Lambda )$ is the renormalization constant associated with the Lagrangian's mass term.

The rainbow-ladder (RL) truncation of the integral equation~\eqref{QuarkDSE} and of the BSE kernel has proven to be a robust and successful symmetry-preserving approximation
of the full tower of equations in QCD when it comes to the description of light ground-state mesons in the isospin-nonzero pseudoscalar and in the vector channels as well as of the 
$N$, $N^*$ and $\Delta$ baryons. The rainbow truncation of the DSE is given by the prescription, 
\begin{equation}
\hspace*{-1mm}
 Z_1^f g^2  D_{\mu\nu} (q) \Gamma_{\nu, f} (k, p) = \big ( Z^f_2 \big)^{\!2}\, \mathcal{G} (q^2) D_{\mu\nu}^\mathrm{free} (q) \frac{\lambda^a }{2} \gamma_\nu\, ,
\label{RLtrunc}
\end{equation}
where we work in Landau gauge in which the free gluon propagator is transverse~\cite{Bashir:2012fs,Serna:2018dwk},
\begin{equation}
    D_{\mu\nu}^\mathrm{free} (q) :=  \delta^{a b}\left(\delta_{\mu \nu}-\frac{q_{\mu} q_{\nu}}{q^{2}}\right ) \! \frac{1}{q^2} \ ,
 \label{freegluon}   
\end{equation}    
and  $\mathcal{G}(q^2)$ in an effective interaction model of the gluon and vertex dressing. In essence, the complexity of the nonperturbative quark-gluon vertex is reduced 
to the one leading Dirac term, where an Abelianized Ward-Green-Takahashi identity,
\begin{equation}
   i q \cdot \Gamma_f (k, p)  =  S^{-1}_f(k) - S^{-1}_f (p) \ , 
 \label{WTIgauge}  
\end{equation}
is enforced which leads to $Z_1^f = Z_2^f$~\cite{Bashir:2012fs}. In perturbation theory this is tantamount to neglecting the contributions of the three-gluon vertex to
$\Gamma_{\mu}(k, p)$ and obviously a drastic simplification of the Slavnov-Taylor identity for the quark-gluon vertex, as it implies equality of the renormalization 
constants of the ghost-gluon vertex and ghost wave function: $\tilde Z_1 = \tilde Z_3$~\cite{Davydychev:2000rt,Alkofer:2008tt,Rojas:2013tza,Aguilar:2018epe,
Aguilar:2018csq,Oliveira:2020yac,Albino:2018ncl}. Furthermore, with the ansatz~\eqref{RLtrunc} we introduce an additional factor $Z_2^f$ to ensure multiplicative 
renormalizability of Eq.~\eqref{QuarkDSE} and thus the renormalization-point independence~\cite{Bloch:2002eq} of the mass function  $M_f (p^2)$:
\begin{equation}
  \Gamma_{\mu}(k, p)=Z_2^f  \gamma_{\mu} \ .
 \label{vertexansatz} 
\end{equation}
With this, the constants, $Z_2^f(\mu,\Lambda)$ and $Z_4^f(\mu,\Lambda)$, are determined using Eqs.~\eqref{EQ:Amu_ren} and \eqref{massmu_ren}, respectively,
which leads to a non-linear coupled renormalization condition~\cite{Hilger:2017jti}.  Their values at $\mu = 2$~GeV are listed in Tab.~\ref{tab:parameters}.

%%%%%%%%%%%%%%%%%%%%%%%%%%%%%%%%%%%%%%%%%%%%%%%%%%%%%%%%%%%%%%%%%%%%%%%%%%%%%%%%%%%%%%%%%%
\begin{figure}[t!]
\centering
  \includegraphics[scale=0.72,angle=0]{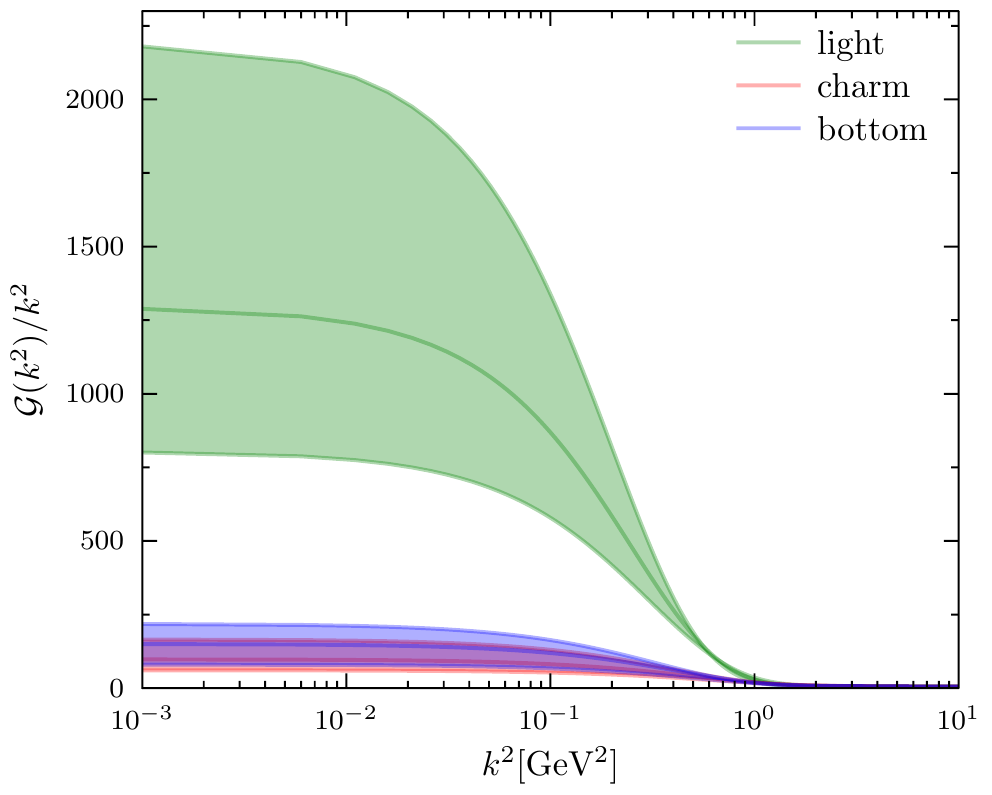}   
   \caption{\label{fig:gluon-inter}  The interaction function $\mathcal{ G}(k^2)/k^2$ in Eq.~\eqref{IR+UV} for different quark flavors and with $\omega_f$ and $\kappa_f$
                                                   reported in Tab.~\ref{tab:parameters}. The shaded bands describe the model uncertainty in varying the interaction parameter $\omega_f$; 
                                                   see Section~\ref{numerics} for details.} \vspace*{-3mm}
\end{figure}
%%%%%%%%%%%%%%%%%%%%%%%%%%%%%%%%%%%%%%%%%%%%%%%%%%%%%%%%%%%%%%%%%%%%%%%%%%%%%%%%%%%%%%%%%%

The ansatz in Eq.~\eqref{RLtrunc} implies that a single dressing function $\mathcal{G}(q^2)$ describes the joint effects of vertex and gluon dressing in the DSE. While this truncation 
is effective and successful for light hadrons for reasons elucidated in Ref.~\cite{Bender:1996bb}, the dynamics in open-flavor mesons is dominated by a wide array of energy scales. 
In particular, the nonperturbative interactions of a light quark and a charm or bottom quark with a gluon cannot be assumed to be similar and therefore be described by
equal dressing functions. And whilst the truncation of Eq.~\eqref{RLtrunc} preserves the axialvector Ward-Green-Takahashi identity (WTI) and therefore chiral symmetry, it 
is clear that the identity~\eqref{WTIgauge} for a bare vertex,
\begin{align} 
  i q \cdot  \gamma  & =   \,   i  k \cdot  \gamma \, Z_f^{-1} (k^2 )   -  i  p \cdot  \gamma  \, Z_f^{-1} (p^2 )  \nonumber  \\
                                & + \, M_f ( k^2 ) Z_f ^{-1}(k^2 )   -  M_f ( p^2 ) Z_f^{-1} (p^2 )  \ ,
\end{align} 
can only hold approximately when $M_f ( k^2 ) \simeq M_f ( p^2) $ and $Z_f ( k^2 ) \simeq Z_f ( p^2)\simeq 1 $  $\forall \, k^2, p^2 > 0$.  This is, at best, the case for very 
heavy quarks as  dynamical chiral symmetry breaking (DCSB) contributes little to their mass functions. Yet, this is far from true for the charm 
quark~\cite{ElBennich:2012tp} and for lighter quarks dressing effects are important. 

In order to introduce the flavor asymmetry in the anti-quark-quark interaction kernel of the $D$ and $B$ mesons, we assume an explicit flavor dependence in the interaction 
$\mathcal{G}_f (q^2)$  and denote it by a subscript. Herein, we will use $\mathcal{G}_u (q^2) = \mathcal{G}_d (q^2) = \mathcal{G}_s (q^2) \neq \mathcal{G}_c (q^2) \neq \mathcal{G}_b (q^2)$.
The dressing function $\mathcal{G}_f(q^2)$  is modeled  after Ref.~\cite{Qin:2011dd}  and consists of the sum of an ansatz in the infrared region, which dominates for  
$|k| < \Lambda_\mathrm{QCD}$  and is suppressed at large momenta, and a second term that implements the regular continuation of the perturbative QCD coupling and 
dominates large momenta,
\begin{equation}
    \frac{\mathcal{G}_f(q^2)}{q^2}  = \,  \mathcal{G}_f^\mathrm{ IR} (q^2) +  4\pi \tilde\alpha_\mathrm{PT} (q^2)  \ ,
\label{IR+UV}    
\end{equation}
where we deliberately absorb a factor $1/q^2$ from the gluon propagator~\eqref{freegluon} in the definition. The expressions for both terms are given by,
\begin{align}
\label{DSEflavor}
 \mathcal{G}_f^\mathrm{IR} (q^2)  &=   \frac{8\pi^2}{\omega^4_f}  D_f\,  e^{-q^2/\omega^2_f}    \nonumber \\   
  4\pi \tilde\alpha_\mathrm{PT}(q^2) &=  \frac{8\pi^2  \gamma_m \mathcal{F}(q^2)}{ \ln \left [  \tau +\left (1 + q^2/\Lambda^2_\textrm{\tiny QCD} \right )^{\!2} \right ] }  \ , 
\end{align}
with $\gamma_m=12/(33-2N_f)$ being the anomalous dimension and $N_f $ the active flavor number, $\Lambda_\textrm{\tiny QCD}=0.234$~GeV, $\tau=e^2-1$, $\mathcal{F}(q^2)=[1-\exp(-q^2/4m^2_t)]/q^2$ and $m_t=0.5$~GeV.  

The ansatz in Eq.~\eqref{DSEflavor} can be parametrized by 
\begin{equation}
 \frac{ \mathcal{G}(q^2)}{q^2 } =\frac{4 \pi \alpha_s (q^2 )}{ q^2+m_g^2 (q^2 )} \ , 
     \quad m_g^2 ( q^2 )=\frac{M_g^4}{q^2 + M_g^2 } \ ,
\end{equation}
where $m_g^2 ( q^2 )$ is an effective gluon mass that vanishes in the ultraviolet and $M_g$ is a mass scale~\cite{Aguilar:2009nf}. The low-momentum component leads 
to an infrared massive and finite interaction consistent with modern DSE and lattice-QCD results and is responsible for DCSB. 

Hence, the flavor dependence is explicit in  the infrared component of the interaction via the constant,
\begin{equation}
   D_f \omega_f : = \kappa_f^3 \ , 
\end{equation}   
where we use equal values of $\kappa_f$ and $\omega_f $ for $f = u,d,s$ and different ones for $ \kappa_c, \omega_c$ and $ \kappa_b,\omega_b$; 
note that $ \kappa_f$ is in unit of GeV. For comparison, we plot $\mathcal{G}_f(q^2) / q^2$ in Fig.~\ref{fig:gluon-inter}, from which it is clear that the 
interaction is strongly attenuated in the heavy sector and the interaction probes more the light quarks in the infrared domain. The interaction strengths
of the heavy quarks overlap within the uncertainty bands due to $\Delta \omega_f$ discussed in Section~\ref{numerics}, albeit that of the charm quark is 
slightly more suppressed contrary to expectation. Indeed, $\mathcal{ G}_b(q^2)/q^2$ can be made weaker than  $\mathcal{ G}_c(q^2)/q^2$ with readjustments 
of  $\omega_b$ and $\kappa_b$, while keeping the mass spectrum of the $B$, $B_s$, $B_c$ and $\eta_b$ virtually the same. However, the decay constants 
of the $B$ mesons suffer a decrease of $20-30$\%.

With this interaction ansatz, we solve the DSE for each quark flavor at space-like momenta, $p^2 >0$, using the parameters reported in Tab.~\ref{tab:parameters}. These 
parameters have been chosen so to reproduce the masses and decay constants studied herein. More precisely, $m_u(\mu) = m_d(\mu)$ and $\kappa_u$ and $\omega_u$  are 
set with the pion's mass and decay constant, $m_s (\mu)$ is fixed likewise with the kaon using $\kappa_s = \kappa_u$, $\omega_s = \omega_u$, and these 
same light-quark parameters are employed in the heavier mesons. Similarly, the values of $m_c (\mu), \kappa_c$ and $\omega_c $ are chosen so to reproduce $m_D$ 
and $f_D$, whereas $m_{\eta_b}$ and $f_{\eta_b}$ are obtained from adjusting $m_b(\mu)$, $\kappa_b$ and  $\omega_b$. We fix the quark masses at $\mu = 19$~GeV
and then evolve them to a scale of 2~GeV at which we compute the quark propagators in the complex plane, as will be discussed shortly in Section~\ref{numerics}.

%%%%%%%%%%%%%%%%%%%%%%%%%%%%%%%%%%%%%%%%%%%%%%%%%%%%%%%%%%%%%%%%%%%%%%%%%%%%%%%%%%%%%%%%%%
\begin{table}[t!]
\begin{center}
\begin{tabular}{ C{5mm} |  C{8mm}| C{7mm} | C{6mm} | C{6mm}| C{7mm} | C{6mm}  |  C{6mm} }
\hline \hline
  $f$  &  $m^{\mu_{19}}_f$  & $m^{\mu_2}_f$  &   $\omega_f $ &  $\kappa_f$ &  $M^E_f$ &  $Z_2^f$  &  $Z_4^f$    \\  
\hline
$u,d$  & 0.0034 & 0.018 & 0.50 & 0.80 & 0.408  & 0.82 & 0.13  \\
$s$      &  0.082 & 0.166 & 0.50 & 0.80 & 0.562  & 0.82 & 0.32   \\
$c$      &  0.903 & 1.272 & 0.70 & 0.60 & 1.342  & 0.94 & 0.53   \\
$b$      &  3.741 & 4.370 & 0.64 & 0.56 & 4.259  & 0.97 & 0.62   \\
\hline \hline
\end{tabular}
\end{center} \vspace*{-1mm}
\caption{\label{tab:parameters} Model parameters [in GeV]: $m^{\mu_{19}}_f = m_f (19\, \textrm{GeV} )$, $m^{\mu_2}_f = m_f (2\,\textrm{GeV} )$,  $\omega_f$ and 
$\kappa_f =(\omega_f D_f)^{1/3}$.  $M^E_f$ is the Euclidean constituent quark mass: $M^E_f = \{p^2|p^2=M^2(p^2)\} $. Renormalization constants in DSE: $Z_2^f$ and 
$Z_4^f$ at $\mu=2$~GeV.  }  \vspace*{-3mm}
\end{table}
%%%%%%%%%%%%%%%%%%%%%%%%%%%%%%%%%%%%%%%%%%%%%%%%%%%%%%%%%%%%%%%%%%%%%%%%%%%%%%%%%%%%%%%%%%

%%%%%%%%%%%%%%%%%%%%%%%%%%%%%%%%%%%%%%%%%%%%%%%%%%%%%%%%%%%%%%%%%%%%%%%%%%%%%%%%%%%%%%%%%%

\subsection{Bound-State Equation}

The axialvector WTI is crucial to satisfy the chiral properties of the Goldstone bosons of QCD and to guarantee that the pion is massless in the chiral limit. The identity 
is derived from chiral transformations and reads~\cite{Itzykson:1980rh}, 
\begin{eqnarray}
  P_{\mu} \Gamma_{5 \mu}^{f g}(k ; P)  &=  &  S_f^{-1} \left (  k_\eta \right) i \gamma_5 +  i \gamma_5 S_g^{-1}\left(k_{\bar \eta}  \right  ) \nonumber  \\ [0.1true cm]   
                                                                & &  - i  \left[ m_f+m_g \right ] \Gamma_5^{f g}(k ; P) \ ,
 \label{axWTI}                                                               
\end{eqnarray}
where  $\Gamma_{5 \mu}^{f g}(k ; P)$ and $ \Gamma_5^{f g}(k ; P)$ respectively denote the axialvector and pseudoscalar vertices for two quark flavors, $f$ and $g$,
and $P$ is the total four-momentum of the meson, $P^2 = -m_M^2$. The short notation for the quark momenta, $k_\eta = k + \eta P$ and $k_{\bar \eta} = k - \bar \eta P$, defines 
momentum-fraction  parameters, $\eta +\bar \eta = 1$, $\eta \in [0,1]$.

In order to constrain the Bethe-Salpeter kernel by the quark propagators $S_f ( k)$, the interaction and the ansatz for the quark-gluon vertex~\eqref{vertexansatz}, one inserts the
DSE~\eqref{QuarkDSE} as well as the  axialvector and pseudoscalar vertices given by,
\begin{align}
  \Gamma_{5 \mu}^{f g}  (k ; P)  =    & \  Z_2^f   \gamma_5 \gamma_{\mu}   \nonumber \\ 
                                            +  \int^\Lambda \! \!   \frac{d^4q}{(2\pi)^4}&  \,  K_{f g} (q, k ; P) \, S_f ( q_\eta ) \Gamma_{5 \mu}^{f g}(q ; P) S_g (q_{\bar \eta} )  \,  ,  
\end{align}
\begin{align}                              
  \Gamma_5^{f g}  (k ; P)   =   &  \  Z_4^f   \gamma_5 \nonumber \\  
                                          +   \int^\Lambda  \!\!  \frac{d^4q}{(2\pi)^4} & \, K_{f g} (q, k ; P) \,  S_f ( q_\eta ) \Gamma_5^{f g} (q ; P) S_g (q_{\bar \eta} )  \, ,                                   
\end{align}
in which $K_{f g}(q, k ; P)$ is the fully-amputated quark-anti-quark scattering kernel and the Dirac- and color-matrix indices are implicit, into Eq.~\eqref{axWTI} 
which leads to the relation ($l=k-q$), 
\begin{eqnarray}
\label{WTImod} 
\lefteqn{\hspace*{-6mm}
  \int^\Lambda \! \!   \frac{d^4q}{(2\pi)^4}   \,  K_{f g}(q, k ; P)   \big [ S_f  (q_{\eta} )  \gamma_5 +  \gamma_5  S_g  (q_{\bar \eta }) \big  ]  = } \nonumber  \\   [0.1true cm]   
     &  &      \hspace*{-1cm}  -  \int^\Lambda \! \! \frac{d^4q}{(2\pi)^4} \,
      \gamma_{\mu} \big [ \Delta_{\mu \nu}^f (l)    S_f  (q_{\eta} )  \gamma_5   +   \gamma_5  \Delta_{\mu \nu}^{g}(l)  S_g (q_{\bar \eta } )  \big ] \gamma_{\nu} \,  ,
\end{eqnarray}
where we define:
\begin{equation}
  \Delta_{\mu \nu}^f (l) = \frac{4}{3}\, \big (Z_2^f \big )^{\! 2}  \, \mathcal{G}_f (l^2)  \left ( \delta_{\mu \nu} - \frac{l_\mu l_\nu}{l^2}\right ) \frac{1}{l^2} \ .
\end{equation} 
Closely inspecting both sides of Eq.~\eqref{WTImod} one realizes that, in a RL truncation with a flavor-dependent interaction, the kernel  $ K_{f g}(q, k ; P) $ on the 
left-hand side must express an average  of the interactions. Note that in the limit $\Delta_{\mu \nu}^f (l) =  \Delta_{\mu \nu}^g (l)$ the identity Eq.~\eqref{WTImod} is satisfied 
by the usual RL kernel,
\begin{equation}
  K(q,k;P) =  - Z_2^2 \, \mathcal{G}\left(l^{2}\right)  D_{\mu \nu}^{\mathrm{free}}(l)\,  \gamma_{\mu} \frac{\lambda^a}{2} \gamma_{\nu} \frac{\lambda^a}{2} \ .
\end{equation}

In a consistent ansatz for $K^{f g}(q, k ; P)$ that satisfies Eq.~\eqref{WTImod}, it can be shown~\cite{Qin:2019oar} that the kernel behaves for large momenta $q \to \infty $ as, 
\begin{equation}
   K_{fg}  \sim   -\, \gamma_\mu  \left(  \frac{\Delta_{\mu \nu}^f + \Delta_{\mu \nu}^g}{2}  \right ) \gamma_\nu  \ ,
\end{equation}
whereas in the infrared limit this becomes,
\begin{equation}
  K_{fg}    \sim   -\, \gamma_\mu  \left (  \frac{\Delta_{\mu \nu}^f \sigma_s^f (0) + \Delta_{\mu \nu}^g \sigma_s^g (0) }{\sigma_s^f (0) + \sigma_s^g (0) }  \right )    \gamma_\nu \, .
\end{equation}
In both cases the kernel tends to an average of interaction functions, in the latter case weighted with flavored quark-dressing functions.

In this light, we choose the ansatz for the kernel, 
\begin{equation}
  \label{RLkernel}
    K_{fg}(k,q;P)   = -  \mathcal{Z}_2^2  \  \frac{\mathcal{G}_{fg}  (l^2)}{l^2 } \frac{\lambda^a }{2} \gamma_\nu \frac{\lambda^a }{2} \gamma_\nu \ ,
\end{equation}
combining the wave-function renormalization constant of both quarks $ \mathcal{Z}_2 (\mu,\Lambda) =\surd Z_2^f\surd Z_2^g$  and introducing the ansatz,
\begin{equation}
    \frac{ \mathcal{G}_{fg}  (l^2) }{ l^2 }= \mathcal{G}_{fg}^\mathrm{ IR}(l^2) +  4\pi \tilde\alpha_\mathrm{PT}(l^2) ,
\end{equation}
where the averaged interaction in the low-momentum domain is described by: 
\begin{equation}
\label{BSEflavor}
 \mathcal{G}_{fg}^\mathrm{IR} (l^2)  =   \frac{8\pi^2}{(\omega_f\omega_g)^2} \sqrt{D_f\,D_g }\,e^{-l^2/(\omega_f\omega_g)} \ ,
\end{equation}

We insert this ansatz in the homogeneous BSE,
\begin{equation}
\label{BSE}
  \Gamma^{fg}_M (k,P) \!=\! \int^\Lambda\! \! \!\frac{d^4q}{(2\pi)^4 } K_{fg} (k,q;P)  S_f (q_\eta) \Gamma^{fg}_M (q,P) S_g(q_{\bar \eta} )  ,
\end{equation}
and obtain Poincaré-invariant solutions which are the \linebreak BSAs, $\Gamma^{fg}_M (k,P)$, in the pseudoscalar channel $J^{PC} = 0^{-+}$. They can be expanded 
in a non-orthogonal base with respect to the Dirac trace: 
\begin{align}
   \Gamma^{fg}_M (k,P)   =   \gamma_5   & \Big [ i E^{fg}_M (k,P) +  \gamma \cdot P\,  F^{fg}_M (k,P)   \nonumber \\  
                                        +\    \gamma \cdot k\;  k \cdot P \,  &  G^{fg}_M (k,P)  +   \sigma_{\mu\nu} k_\mu P_\nu \, H^{fg}_M (k,P) \Big ] \, .
 \label{PS-BSA}                                      
\end{align}     
We remind, though, that mesons with unequal quarks, such as the kaon, $D$ and $B$ mesons, are not eigenstates of the charge-conjugation operator
defined as,
\begin{equation}
   \Gamma_M (k,P) \stackrel{C}{\longrightarrow} \bar{\Gamma}_M (k,P) :=C \Gamma^{T}_M (-k,P) C^{T} \ .
\end{equation}
Thus, $\bar \Gamma_M (k,P)  = \lambda_c \Gamma_M (k,P)$ does not imply $\lambda_c =\pm 1$  for their charge parity. For mesons made of valence quarks 
with equal current mass and $J^{PC} = 0^{-+}$, the constraint that the Dirac base in~\eqref{PS-BSA} satisfies $\lambda_c  = +1$ requires the scalar amplitudes  
to be even under $k\cdot P \rightarrow   -k\cdot P$. Each amplitude, $\mathcal{ F}_i= E_M$, $ F_M$, $G_M$, $H_M $, can furthermore be decomposed in 
terms of,
\begin{equation}
\label{BSA-kaon-D}
   \mathcal{F}_i (k,P) = \mathcal{ F}^0_i (k,P) + k\cdot P\, \mathcal{F}^1_i (k,P),  
\end{equation}
in which $\mathcal{F}^{0,1}_i (k,P)$ are even under $k\cdot P \rightarrow   -k\cdot P$. As a consequence, the neutral pion and quarkonia have the property 
$\mathcal{F}_1(k,P) \equiv  0$, yet $\mathcal{F}_1(k,P)$ will contribute to flavored mesons.

%%%%%%%%%%%%%%%%%%%%%%%%%%%%%%%%%%%%%%%%%%%%%%%%%%%%%%%%%%%%%%%%%%%%%%%%%%%%%%%%%%%%%%%%%%
\begin{figure}[t!]
\begin{center}
  \includegraphics[scale=0.4]{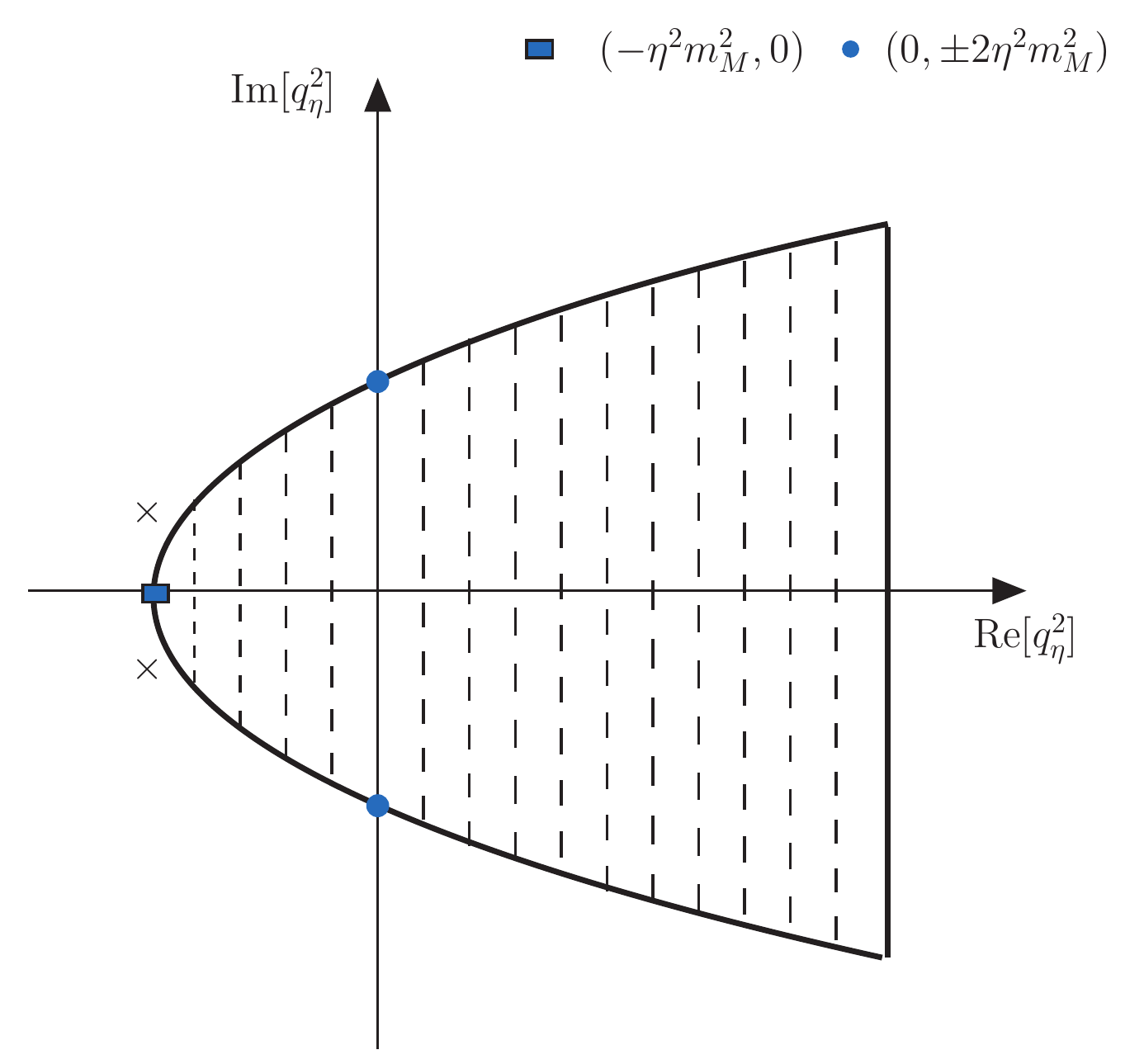} 
\end{center}\vspace*{-1mm}
\caption{The integration domain defined by the BSE for a meson of mass $m_M$ described by a parabola in the complex $q^2_\eta$-plane.}
\label{parabola}
\vspace*{-4mm}
\end{figure}
%%%%%%%%%%%%%%%%%%%%%%%%%%%%%%%%%%%%%%%%%%%%%%%%%%%%%%%%%%%%%%%%%%%%%%%%%%%%%%%%%%%%%%%%%%

We apply the Nakanishi normalization condition~\cite{Nakanishi:1965zz,Nakanishi:1965zza} which makes use of the eigenvalue trajectory $\lambda (P^2)$
of the BSE,
\begin{eqnarray}
  \left(\frac{\partial \ln (\lambda)}{\partial P^{2}}\right)^{-1}  & = & \ \operatorname{tr}_{CD}  \int \frac{d^4k}{(2\pi)^4}  \, \bar{\Gamma}^{fg}_M (k ;-P) 
  \nonumber \\   [0.2true cm]   
                      & \times &  S_f  (k_{\eta } ) \Gamma^{fg}_M  (k ; P) S_g (k_{\bar \eta })  \, , 
\label{nakanishinorm}                      
\end{eqnarray} 
to normalize the BSA and verify the normalization with the usual canonical method:
\begin{eqnarray}
     2 P_{\mu}   & =  &  \frac{\partial}{\partial P_{\mu}}   \int \frac{d^4k}{(2\pi)^4}  \operatorname{Tr}_{CD} \big  [ \Gamma(k ;-K)  \nonumber  \\   [0.1true cm]   
                       & \times & S (k_\eta ) \Gamma(k ; K) S (k_{\bar \eta} ) \big ]  \Big  |_{P^2=K^2 = -M^2 } \ .
\label{canonical}                       
\end{eqnarray}
The normalization  is required to calculate the weak decay constant of the pseudoscalar meson:
\begin{equation}
\label{fdecay} 
  f_M  P_\mu = \frac{N_c \mathcal{Z}_2}{\sqrt{2} }\int^\Lambda\!  \frac{d^4k}{(2\pi)^4} \,\operatorname{Tr}_D \left [ \gamma_5\gamma_\mu\,  \chi_M (k_\eta,k_{\bar \eta}) \right ] \, .
\end{equation}
Henceforth, $\chi_M (k_\eta,k_{\bar \eta})  := S_f (k_\eta) \Gamma_M^{fg} (k, P) S_g (k_{\bar \eta} )$  defines the Bethe-Salpeter wave function. As already noted, 
the quark momenta, $k_\eta = k + \eta P$ and $k_{\bar \eta} = k - \bar \eta P$,  define momentum-fraction  parameters; no observables can depend on them owing 
to Poincaré covariance.  

The weak decay constant may also be inferred from the Gell-Mann-Oakes-Renner (GMOR) relation which is just a different expression of the axialvector WTI 
that describes the axialvector-current conservation in the chiral limit; see for instance Refs.~\cite{Rojas:2014aka,Chen:2019otg} for details of the calculation.  
Comparing the decay constant obtained with Eq.~\eqref{fdecay}  and with the GMOR relation provides us an additional check of the kernel in Eq.~\eqref{RLkernel}
and we find variations for $f_D$ and $f_B$ of the order of 3~\%.

%
%%%%%%%%%%%%%%%%%%%%%%%%%%%%%%%%%%%%%%%%%%%%%%%%%%%%%%%%%%%%%%%%%%%%%%%%%%%%%%%%%%%%%%%%%%
\begin{figure*}[t!] 
\centering
\hspace*{3mm}
 \begin{minipage}{0.45\textwidth}
  \includegraphics[scale=0.57,angle=0]{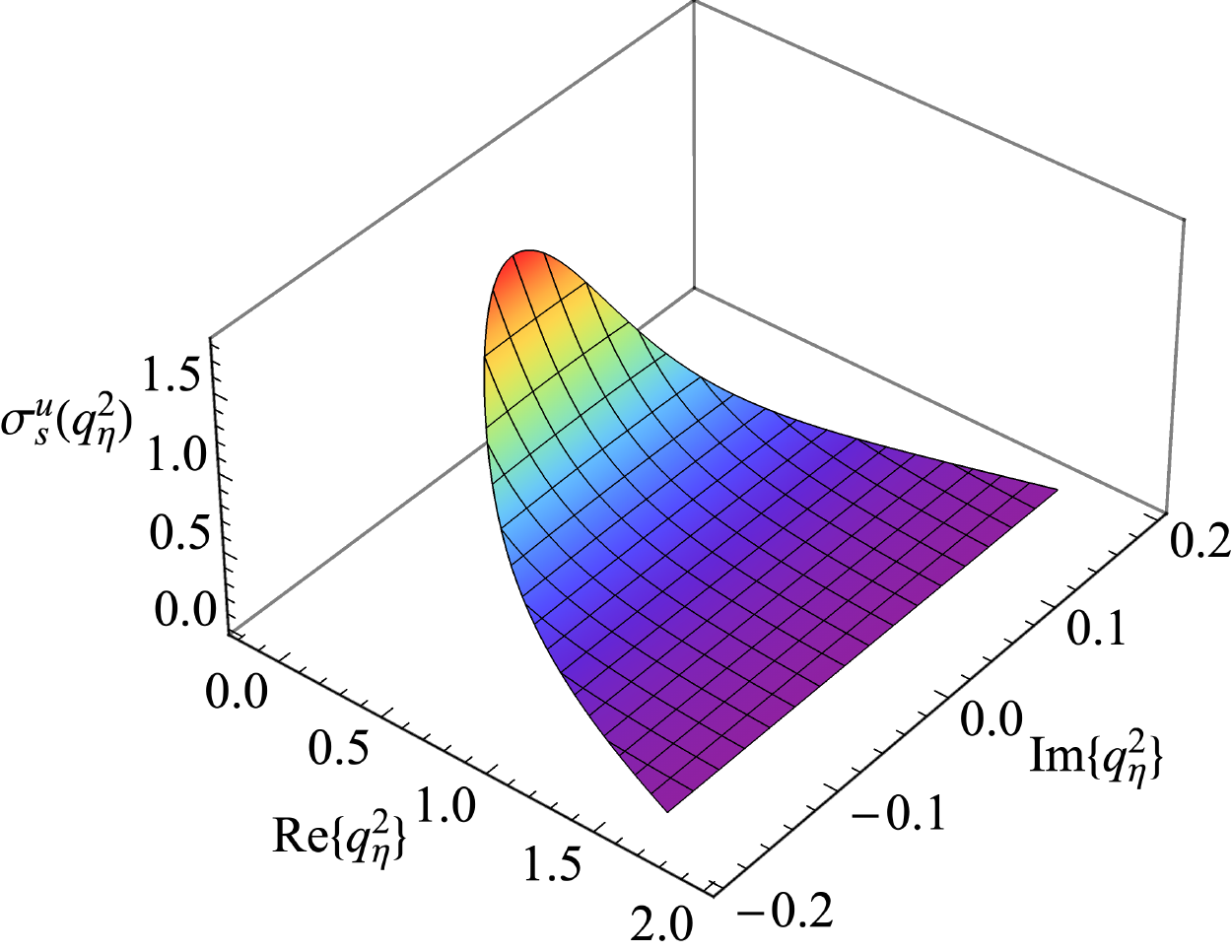}
  \end{minipage}
 \hfill
    \begin{minipage}{0.48\textwidth}
  \includegraphics[scale=0.57,angle=0]{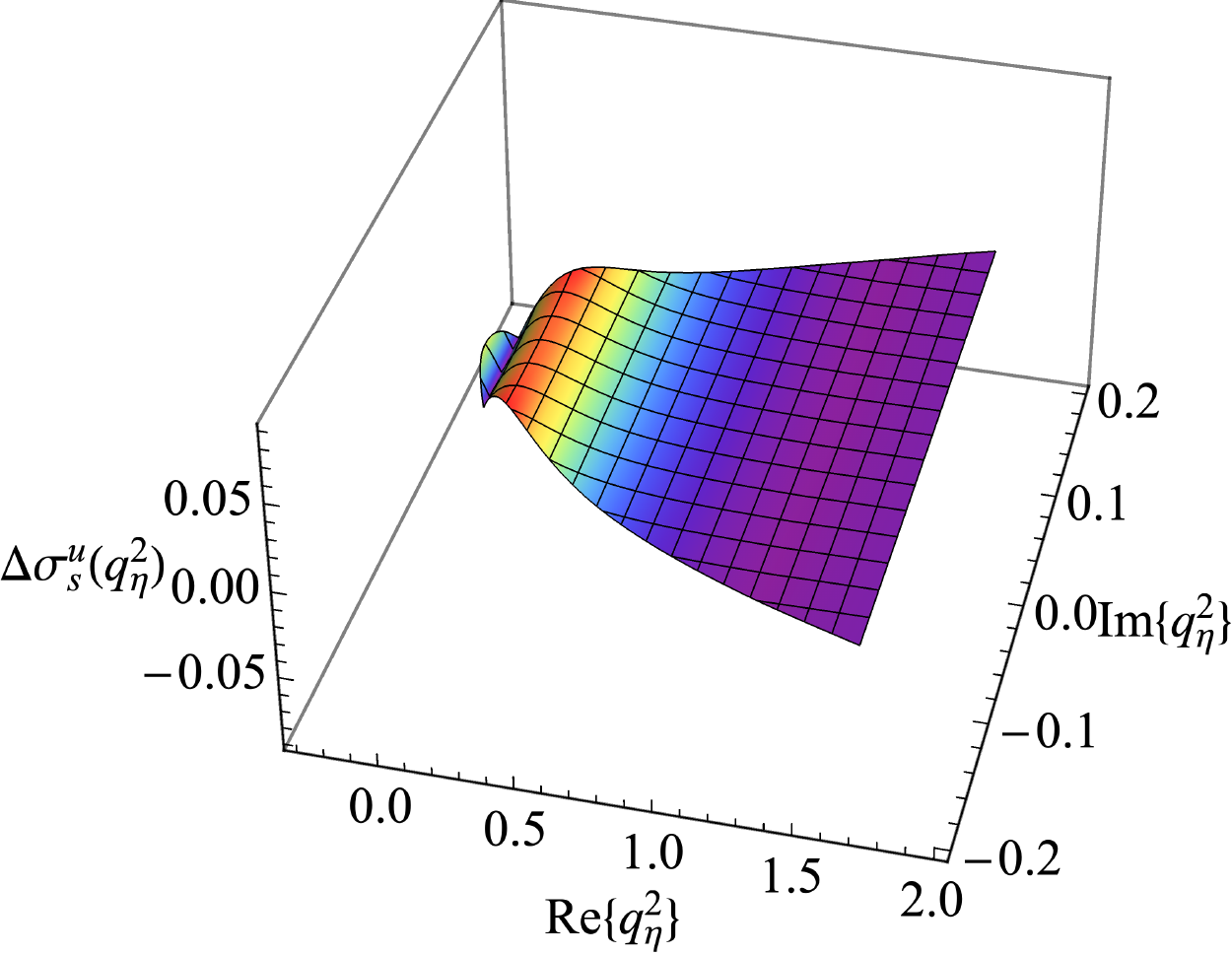}
     \end{minipage}
\vspace*{3mm}
 \caption{\label{fig:deltasigmau}  Comparison of $\sigma_{\rm s}^u(q^2_\eta)$ and $\Delta\sigma_{\rm s}^u(q^2_\eta)$ for the $u$-quark on the complex plane
  ($P^2 = -m_\pi^2$, $\eta =1/2$, $m_\pi=0.140$~GeV). }  
\end{figure*}
%%%%%%%%%%%%%%%%%%%%%%%%%%%%%%%%%%%%%%%%%%%%%%%%%%%%%%%%%%%%%%%%%%%%%%%%%%%%%%%%%%%%%%%%%%
%
\begin{figure*}[ht!] 
\centering
\hspace*{3mm}
 \begin{minipage}{0.45\textwidth}
  \includegraphics[scale=0.57,angle=0]{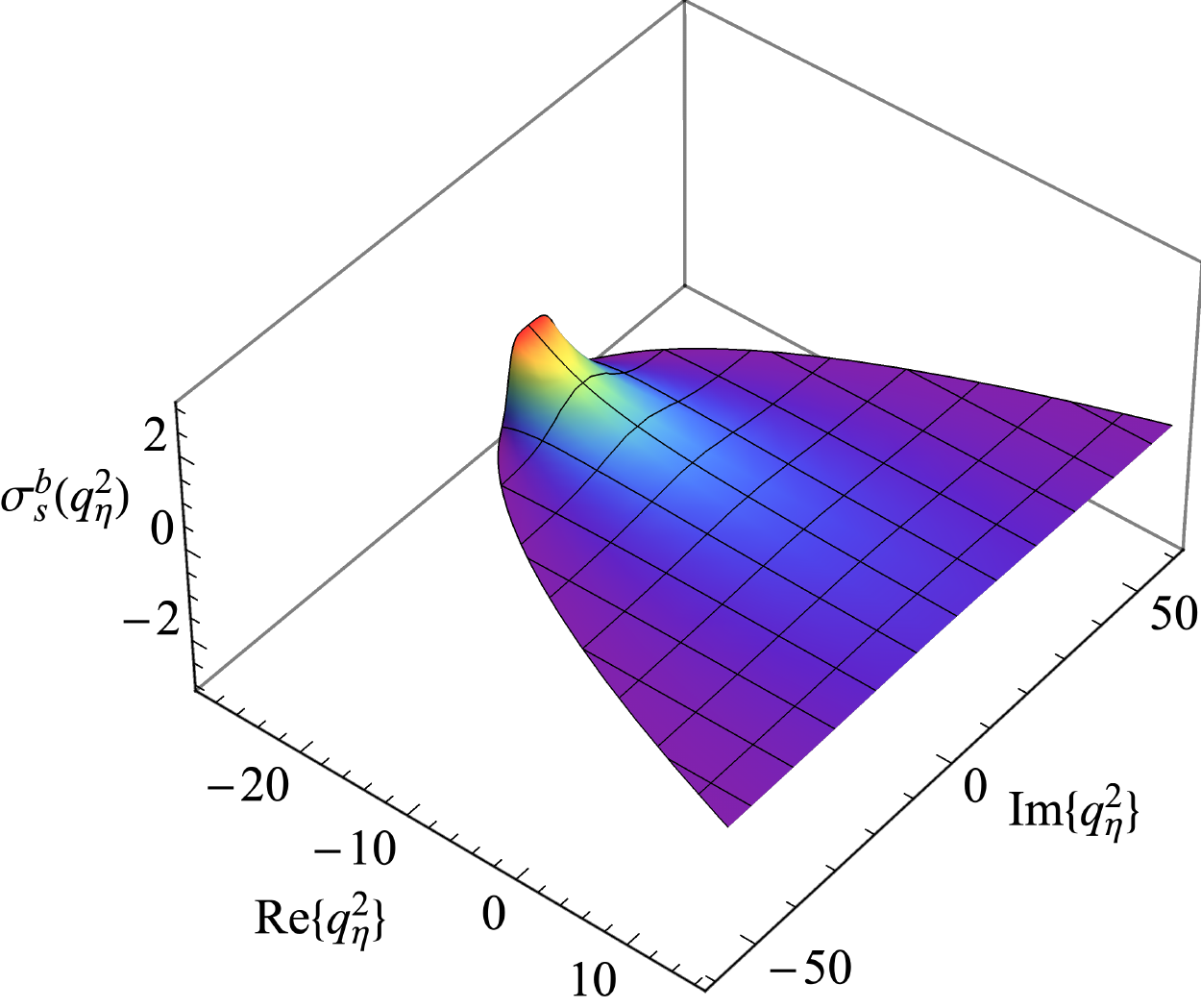}
  \end{minipage}
 \hfill
    \begin{minipage}{0.48\textwidth}
  \includegraphics[scale=0.63,angle=0]{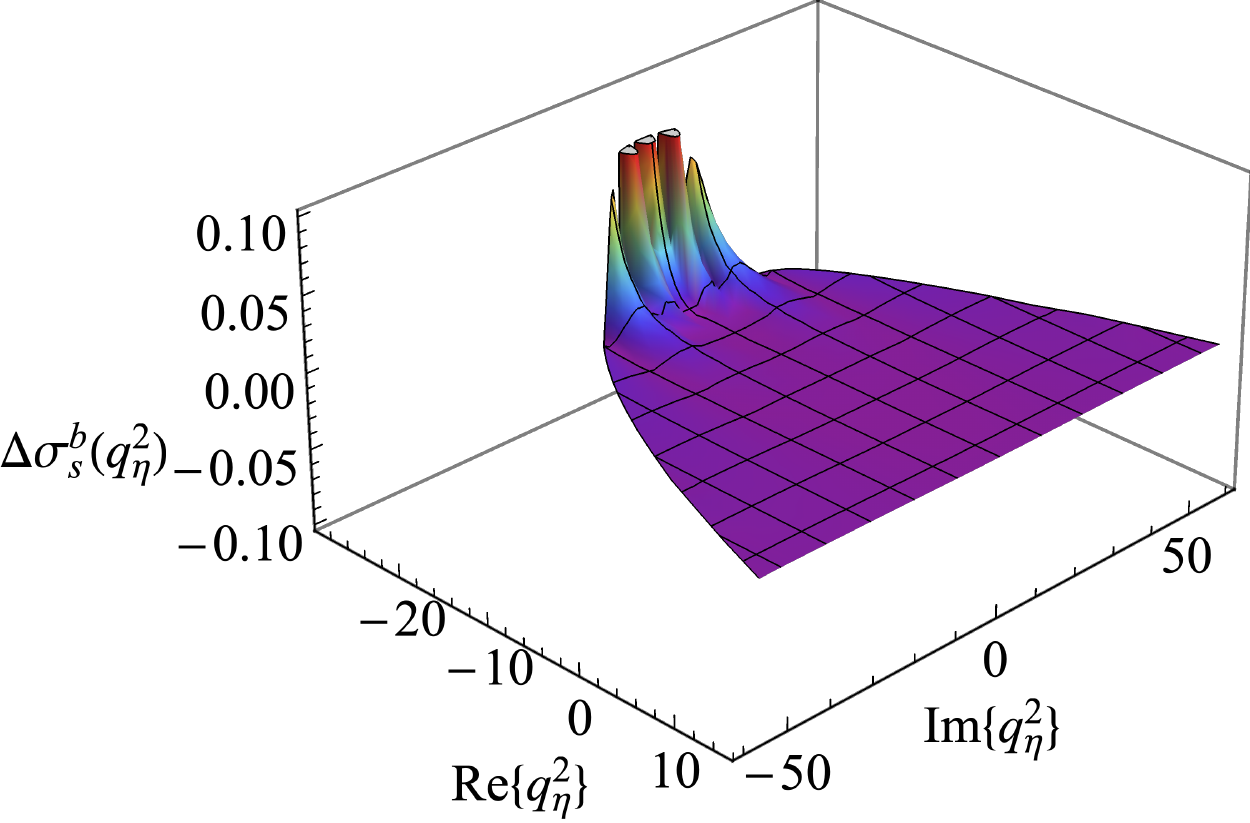}
     \end{minipage}
\vspace*{3mm}
\caption{\label{fig:deltasigmab} Comparison of $\sigma_{\rm s}^b (q^2_\eta)$ and $\Delta\sigma_{\rm s}^b(q^2_\eta)$ for the $b$-quark on the complex plane
     ($P^2 = -m_{\eta_b}^2$, $\eta =1/2$, $m_{\eta_b} = 9.392$~GeV). }  \vspace*{-2mm}
\end{figure*}
%%%%%%%%%%%%%%%%%%%%%%%%%%%%%%%%%%%%%%%%%%%%%%%%%%%%%%%%%%%%%%%%%%%%%%%%%%%%%%%%%%%%%%%%%%

%%%%%%%%%%%%%%%%%%%%%%%%%%%%%%%%%%%%%%%%%%%%%%%%%%%%%%%%%%%%%%%%%%%%%%%%%%%%%%%%%%%%%%%%%%

\subsection{Numerical Results on the Complex Plane  \label{numerics} }

The numerical solution of the BSE~\eqref{BSE} implies the knowledge of the quark propagator,
\begin{eqnarray}
   S_f (q_\eta) & = & -i \gamma \cdot q_\eta \, \sigma_{\rm v}^f (q_\eta^2 )+\sigma_{\rm s}^f (q_\eta^2 ) \nonumber \\ 
                       & = &  Z_f (q_\eta^2 ) / \left [ i \gamma \cdot q_\eta +M_f (q_\eta^2 ) \right  ] \ ,
\end{eqnarray}
and likewise for $ S_f (q_{\bar\eta } )$.  In Euclidean space, the arguments $q_\eta^2$ and $q_{\bar\eta}^2$ define parabolas on the complex plane, \vspace*{-3mm}
\begin{eqnarray}
   q^2_\eta  & = &  q^2  - \eta^2 m^2_M  + 2 i \eta \, m_M  | q | z_q \  , \\  
   q^2_{\bar\eta}  & = &  q^2  - {\bar\eta}^2 m^2_M  - 2 i \bar \eta \, m_M  | q | z_q \  ,
\end{eqnarray}
where $z = q\cdot P /|q||P|$, $-1 \leq z \leq +1$, is an angle. In Fig.~\ref{parabola}, we illustrate a typical situation encountered in numerical studies of the DSE with an external
time-like momentum.

In this figure, the points $(-\eta^2 m^2_M ,0)$ and  $(0,\pm2\eta^2 m_M^2 )$  are respectively the intersection points with the real and imaginary axis defined by,
\begin{eqnarray}
   \text {Re}\left [q^2_\eta\right ]  & = & q^2-\eta^2 m^2_M \, ,\\ 
    \text{Im}\left [q^2_\eta \right]  & = &  2\eta |q| m_Mz_q\,  ,
\end{eqnarray}
and similarly for $\bar \eta$. We note that the size of the parabolic domain is determined by the meson mass $m_M$ and the parabola is fully defined once $m_M$ 
and $\eta$ are known. Since the entire interior of the parabola is sampled in the numerical integration of the BSE, singularities inside this domain must be avoided. 
For the simple case of the pion, where $m_M  < 1$~GeV, these complex-conjugate singularities will be outside the parabolic region, as illustrated with the symbol 
``$\times$" in Fig.~\ref{parabola}\,\footnote{These complex-conjugated singularities are merely illustrative, as there may occur additional singularities and even cuts deeper  
into the complex plane and off the real axis; see Ref.~\cite{El-Bennich:2016qmb}, for example.}. In this case, the propagator functions $\sigma_\mathrm{s,v}^f$ are analytic
and Cauchy's integral theorem can be applied. On the other hand, with increasing meson mass, such as for the $D$ where $m_{D}  > 1$ GeV, singularities may lie 
within the integration domain and  the parameters $\eta$ and $\bar \eta$ can be chosen to adapt the parabola size and shift external momentum from one 
constituent propagator to the other. In doing so, there is a limiting bound-state mass for which the integration domain starts to overlap these singularities.

%%%%%%%%%%%%%%%%%%%%%%%%%%%%%%%%%%%%%%%%%%%%%%%%%%%%%%%%%%%%%%%%%%%%%%%%%%%%%%%%%%%%%%%%%%

\begin{table*}[t!]
\centering
\begin{tabular}{c|c|c|c|c}
  \hline \hline
  $f$& $z_1$& $m_1 $&$z_2$&$m_2$\\ [0.7mm]
  \hline
  $u$	  & $(0.387,0.243) $ & $(0.503,0.258)$ & $(0.142,-0.002)$ & $(-0.900,0.666)$   \\
  $s$  & $(0.432,0.155)$ &  $(0.667,0.318)$ & $(0.148,-0.024)$ & $(-1.143,0.641)$    \\
  $c$	 & $(0.494,1.040) $ &  $(1.822,0.273)$ & $(0.035,-0.037)$ & $(-3.011,0.000)$       \\
  $b$  & $(0.501,1.213) $ &  $(5.254,0.466)$ & $(0.015,0.013)$ & $(-8.376,-0.019) $    \\
\hline \hline
\end{tabular}
\caption{\label{tab:2ccp}   Parameters of the \texttt{ccp} representation of the propagators~\eqref{ccp} for $N=2$ complex-conjugate poles. The pair $(x, y)$ represents the 
             complex number $x+iy$.}     
\end{table*}

%%%%%%%%%%%%%%%%%%%%%%%%%%%%%%%%%%%%%%%%%%%%%%%%%%%%%%%%%%%%%%%%%%%%%%%%%%%%%%%%%%%%%%%%%%

\begin{table*}[t!]
\centering
\begin{tabular}{ C{3cm} ||  C{1.7cm}| C{1.7cm} | C{1.7cm} || C{1.7cm}| C{1.7cm} | C{1.7cm} }
\hline \hline
Mesons/Observables  & $m_M$ & $m^\textrm{exp.}_M$ & $\epsilon^m_r$  [\%] & $f_M$  &  $f^\textrm{exp./lQCD}_M$ & $\epsilon^f_r$  [\%]   \\ [0.7mm]
\hline
$\pi (u\bar d)$&0.136 &0.140 & 2.90 &  $0.094^{+0.001}_{-0.001}$ & $0.092(1)$  & 2.17  \\  
$K(s\bar u)$  &0.494 & 0.494 & 0.0 & $0.110^{+0.001}_{-0.001}$ & $0.110(2)$ & 0.0   \\   
$D_u(c \bar u) $  & $1.867^{+0.008}_{-0.004}$ &1.870 &0.11 &  $0.144^{+0.001}_{-0.001}$ & $0.150(0.5)$ &4.00   \\
$D_s(c \bar s) $  &  $2.015^{+0.021}_{-0.018}$  &1.968 &2.39 & $0.179^{+0.004}_{-0.003}$ & $0.177(0.4)$&1.13  \\
 $\eta_c(c \bar c) $  &  $3.012^{+0.003}_{-0.039}$ & 2.984 &0.94  & $0.270^{+0.002}_{-0.005}$ & 0.279(17) & 3.23 \\
 $\eta_b(b \bar b) $  & $9.392^{+0.005}_{-0.004}$ & 9.398 &0.06  &  $0.491^{+0.009}_{-0.009}$ & 0.472(4) & 4.03   \\
\hline \hline
\end{tabular}
\caption{\label{tab:psproperties} Masses and decay constants [in GeV] of pseudoscalar mesons. Experimental masses and leptonic decay constants
are taken from the Particle Data Group~\cite{PDG} except for the $D$ and $D_s$ decay constants which are FLAC 2019 averages~\cite{Aoki:2019cca}
and $f_{\eta_c}$ which is from Ref.~\cite{McNeile:2012qf}.  The relative deviations from experimental values are  given by $\epsilon^v_r  =100\% \, | v^\textrm{exp.} 
- v^\textrm{th.} | / v^\textrm{exp.}$.}     
\end{table*}   

%%%%%%%%%%%%%%%%%%%%%%%%%%%%%%%%%%%%%%%%%%%%%%%%%%%%%%%%%%%%%%%%%%%%%%%%%%%%%%%%%%%%%%%%%%

\begin{table*}[h!]
\centering
\begin{tabular}{ C{3cm}||  C{1.7cm}| C{1.7cm} | C{1.7cm} || C{1.7cm}| C{1.7cm} | C{1.7cm} }
\hline \hline
Mesons/Observables  & $m_M$ & $m^\textrm{exp.}_M$ & $\epsilon^m_r$  [\%] & $f_M$  &  $f^\textrm{lQCD}_M$ & $\epsilon^f_r$  [\%]   \\ [0.7mm]
\hline
$B_u(b\bar u)$  & $5.277^{+0.008}_{-0.005} $ & 5.279 & 0.04 & $0.132^{+0.004}_{-0.002}$ & $0.134(1)$ &4.35    \\
$B_s(b\bar s)$  & $5.383^{+0.037}_{-0.039}$  & 5.367 & 0.30 & $0.128^{+0.002}_{-0.003}$ &  $0.162(1)$ & 20.50 \\
$B_c(b\bar c)$  & $6.282^{+0.020}_{-0.024}$  & 6.274 & 0.13  & $0.280^{+0.005}_{-0.002}$  & $0.302(2)$ & 7.28 \\
 $\eta_b(b \bar b) $ & $9.383^{+0.005}_{-0.004}$ & 9.398 & 0.16 & $0.520^{+0.009}_{-0.009}$ & $0.472(4) $ &10.17  \\
\hline \hline
\end{tabular}
\caption{\label{tab:Bproperties} Masses and decay constants [in GeV] of the $B$ mesons and $\eta_b$ calculated with the hybrid approach of using a \texttt{2ccp} 
representation for the bottom quark and the numerical \texttt{cp} solutions for the $u$, $s$ and $c$ quarks. Experimental masses are taken from the Particle Data 
Group~\cite{PDG}. The leptonic decay constants of the $B_u$ and $B_s$ are the FLAC 2019 averages~\cite{Aoki:2019cca} and those of the $B_c$ and $\eta_b$ 
are from Ref.~\cite{McNeile:2012qf}. The relative deviations are as in Tab.~\ref{tab:psproperties}.} \vspace*{-2mm}
\end{table*}  

%%%%%%%%%%%%%%%%%%%%%%%%%%%%%%%%%%%%%%%%%%%%%%%%%%%%%%%%%%%%%%%%%%%%%%%%%%%%%%%%%%%%%%%%%%

The numerical application of Cauchy's integral theorem we employ is explained in detail in Ref.~\cite{Krassnigg:2009gd} and our parametrization of the parabola 
contour is described in Ref.~\cite{Rojas:2014aka}. In particular, we use a distribution of momenta on the contour that is skewed towards the vertex of the parabola.
As an example, we plot the real part of $\sigma_s^u (q^2_\eta)$ on the complex plane in the left-hand pannel of Fig.~\ref{fig:deltasigmau}, where the maximal hadron 
mass that can be reached is $m^\textrm{max}_M =0.2~\textrm{GeV} > m_\pi$. Clearly, the dressing function is analytical in this complex domain. 

In case of the $B$ mesons we consider, their large masses do not allow for an optimized $\eta$ and $\bar \eta$ pair that produces a parabola free of singularities 
in the $b$-quark propagator and in the light(er)-quark propagator. We therefore resort to a complex-conjugate pole representation of the $b$-propagator for these 
mesons. The BSA of the $\eta_b$, on the other hand, is obtained with the propagator solution on the complex plane, as the conjugate-complex 
singularities remain outside the parabolic region which can be inferred from the left-hand panel of Fig.~\ref{fig:deltasigmab}. 

Hence in order to compute the static properties of ground-state $B_{u,s,c}$ mesons, we combine two approaches. Due to the issue of unavoidable 
singularities in the $b$-quark propagator on the complex plane, we implement a complex conjugate pole (\texttt{ccp}) parametrization for the heavy-quark 
propagator,  while for the $u$, $s$ and $c$ quarks we use their solutions on the complex plane. The \texttt{ccp} parametrization is given by, 
\begin{equation}
\label{ccp}
 S_f  (q )  =  \sum^N_{k=1}  \left [   \frac{z_k^f}  {i \gamma\cdot q  +  m_k^f  }  +   \frac{\big (z_k^f\big )^{\!*} }  {i \gamma\cdot q  +  \big (m_k^f\big )^{\!*}  } \right  ] \ ,
\end{equation}
$m_k^f$ and $z_k^f$ being complex numbers. These parameters are fitted to the DSE solution~\eqref{DEsol} for $N=2$ on the real space-like axis $p^2\,\in[0,\infty)$, 
and the thus obtained \texttt{ccp} representation is then analytically extended to complex momenta.  Since we make use of the this representation to calculate the 
Mellin moments~\eqref{moment-BSA} of the LCDA, we list their parameters in Tab.~\ref{tab:2ccp} for all flavors. 

Obviously, we want to make sure that these parametrizations constitute a realistic reproduction of the dressing functions on the complex plane. To this end, 
we define the function,
\begin{equation}
   \Delta \sigma_{\rm s}(q^2) = \left |\sigma^\texttt{cp}_{\rm s}(q^2) -  \sigma^\texttt{2ccp}_{\rm s}(q^2) \right | \ ,
\end{equation}
where the superscripts \texttt{cp} and \texttt{2ccp} denote respectively numerical solutions on the complex plane and solutions using Eq.~\eqref{ccp} with two complex 
conjugate poles and the fitted parameters in Tab.~\ref{tab:2ccp}. As can be seen in Figs.~\eqref{fig:deltasigmau} and \eqref{fig:deltasigmab}, the deviations 
$\Delta \sigma_s(q^2 )$ are noticeable near the vertex of the parabola, yet the scale of these variations is dwarfed by the magnitude of $\sigma^\texttt{cp}_s(q^2)$.
We also note that the weak decay constant of the pion calculated with the \texttt{2ccp} approach differs by only 3\% from the \texttt{cp} result; similar observations
hold for the kaon and $D$ mesons and the $\eta_b$ mass is almost equal with either method, while the decay constant differs by 6\%. We conclude that 
the use of  the \texttt{2ccp} bottom-propagator is a reliable approach and gives us confidence to calculate the $B$ meson's static properties.

Our results for the masses and leptonic decay constants of the ground-state pseudoscalar mesons are listed in Tab.~\ref{tab:psproperties}, from which it becomes clear
that they are in very good agreement with experimental data when available or lattice-QCD results otherwise. In this table, we exclude the $B$ mesons, the reason for which 
is that the above mentioned hybrid approach is employed. The results for the $B$ mesons as well as for the $\eta_b$ using the hybrid approach are found in 
Tab.~\ref{tab:Bproperties}. The masses of these mesons are in excellent agreement with experimental values, while our decay constants compare reasonably well 
with simulations of lattice-QCD. 

The theoretical uncertainties are obtained as follows: in adjusting the dressing function of the interaction~\eqref{DSEflavor} in the  light-meson sector, we set the scale 
with the pion and kaon masses and weak decay constants. As well known, these observables are rather insensitive to a range $\omega \pm\Delta\omega$ and we set 
an upper and lower limit, $0.45 \leq \omega_{u,s} \leq 0.55$~GeV, about the central value $\omega_{u,s}=0.5$~GeV, as depicted by the uncertainty bands in Fig.~\ref{fig:gluon-inter}.
Having introduced this uncertainty in the light  sector,  the repercussions are immediate in computing the properties of the $D_u$ and $B_u$; their mass uncertainties 
are due to the low-energy scale insensitivity of $\omega_{u,s}$. Likewise, we observe the sensitivity of $D_u$ to variations of 10\% in $\omega_c$ and this yields and 
error estimate for the $D_s$, $\eta_c$ and  $B_c$. Finally, we fix the bottom quark at the $\eta_b$ mass scale and check the combined effect of permissible variations 
of $\omega_b$ in the BSE of the $B_s$, $B_c$ and $\eta_b$ that ensure the $\eta_b$ mass stays within 1\% of its central mass value. 

We remind that these results are not achieved without the implementation of the flavor dependence of the interaction in Eqs.~\eqref{DSEflavor} and \eqref{BSEflavor}. 
The flavor dependence is important to accommodate the fact that heavy quarks probe shorter distances than the light quarks at the corresponding quark-gluon 
vertices, thereby implying a smaller coupling strength for heavy quarks.

%%%%%%%%%%%%%%%%%%%%%%%%%%%%%%%%%%%%%%%%%%%%%%%%%%%%%%%%%%%%%%%%%%%%%%%%%%%%%%%%%%%%%%%%%%
%%%%%%%%%%%%%%%%%%%%%%%%%%%%%%%%%%%%%%%%%%%%%%%%%%%%%%%%%%%%%%%%%%%%%%%%%%%%%%%%%%%%%%%%%%

\section{Distribution Amplitudes  \label{pda} }

A unique leading-twist LCDA exists for any pseudoscalar meson $M$ with total momentum $P$ and is defined in QCD via a meson-to-vacuum matrix element of a nonlocal 
anti-quark-quark light-ray operator as,
\begin{align}
   \langle 0   | \bar{q}_f & (y_2 n ) \mathcal{W} \left[ y_2n, y_1 n \right  ] \gamma \cdot n\, \gamma_{5}\,  q_g ( y_1 n) | M(P) \rangle  \nonumber \\
   & =  i f_{M} \, n\cdot p \int_{0}^{1} dx\, e^{-i  n \cdot P \left( y_1 x + y_2 \bar x \right ) } \phi_{M} (x , \mu ) \ ,
 \label{LCDAdef}   
\end{align}
where $f_M$ is the weak decay constant of the pseudoscalar meson, $n$ is an auxiliary light-like four-vector with $n^2 =0$, $x=k^z/P^z$ is the momentum fraction of the quark 
in the infinite-momentum frame with $\bar x = 1-x $, $y_1$ and $y_2$ are real numbers, and $\mathcal{W} \left[ y_2n, y_1 n \right  ] $ is a light-like Wilson line connecting the 
quark fields $q_f$ and $q_g$ to produce gauge-invariant quantities for any choice of $y_2$ and $y_1$. The momentum-space distribution amplitude $\phi_{M} (x , \mu )$ is the Fourier 
transformed distribution $\tilde \phi_M(y,\mu)$ in coordinate space. 

In principle, $\phi_{M} (x , \mu )$ is directly accessible from the light-front wave function~\cite{Brodsky:1997de} by integrating over the meson's transverse  
momentum,
\begin{equation}
   f_{M}\, \phi_{M} (x , \mu )  = \frac{1}{(2\pi)^3} \int^{\mu^2 } \!\!  d^2k_\perp \, \psi^{\uparrow \downarrow \pm \downarrow \uparrow}_M (x, k_\perp) 
\end{equation}
where $ \psi_M (x, k_\perp)$ is the Fourier transform of the positive-energy projection of the Bethe-Salpeter wave function evaluated at equal time, 
$y^+ = y^3 + y^0 = 0$, in coordinate space~\cite{Lepage:1980fj}. 

In calculating the BSE in momentum space, however, we work in Euclidean space amenable to numerical calculations. We therefore
do not have direct access to the light-front wave function $\psi_M (x, k_\perp)$. A method to eschew the calculation of $\psi_M (x, k_\perp)$ consists of computing 
instead Mellin moments, \vspace*{-2mm}
\begin{equation}
   \left\langle x^{m}\right \rangle = \int_{0}^{1} dx\,  x^{m} \phi_M(x,\mu)  \ ,
\label{momentadef}   
\end{equation} 
using Eq.~\eqref{LCDAdef} and then reconstructing the LCDA from these moments. In particular, the zeroth moment serves to normalize the distribution amplitude
and we choose,
\begin{equation}
  \langle x^{0}  \rangle =  \int_{0}^{1} d x \, \phi_M (x,\mu )  = 1 \ .
\label{zeromomentum}  
\end{equation}

In order to make use of Eq.~\eqref{LCDAdef}, we need to Fourier transform the matrix element to momentum space where after appropriate use of the LSZ 
reduction formula it can be expressed as the light-front projection of the Bethe-Salpeter wave function $ \chi_M (k_\eta,k_{\bar \eta})$,
\begin{align}
  f_M \phi_M (x , \mu) = & \; \frac{ \mathcal{Z}_2 N_c}{\sqrt{2}} \;\mathrm{Tr}_D\! \int^{\Lambda}  \!\!  \frac{d^4k}{(2\pi)^4} \, \delta  (n \cdot k_\eta - x n\cdot P ) \nonumber \\
                                   &   \times  \gamma_5 \, \gamma \cdot n \ \chi_M (k_\eta,k_{\bar \eta})\, ,
\label{BSA-LCDA}  
\end{align}
with the choice  $n \cdot P = - m_M$ in the rest-frame of the meson. 

With this, one may apply the integral in Eq.~\eqref{momentadef}  to both sides of Eq.~\eqref{BSA-LCDA} which, employing the property of the 
Dirac function $\int_{0}^{1} d x\, x^{m} \delta(a-x b) = \frac{a^{m}}{b^{m+1}} \,\theta(b-a) $, leads to the integral, 
\begin{eqnarray}
  \left  \langle x^{m}\right \rangle   & = &   \frac{\mathcal{Z}_2N_c}{\sqrt{2}f_M} \; \mathrm{Tr}_D \! \int^{\Lambda}  \!\! \frac{d^4k}{(2\pi)^4} 
  \frac{ ( n\cdot k_\eta )^m}{ ( n\cdot P )^{m+1}  }     \nonumber   \\  [0.2true cm]  
  &  &  \times\   \gamma_5 \, \gamma \cdot n \; \chi_M (k_\eta,k_{\bar \eta})\, .
  \label{moment-BSA}          
\end{eqnarray}
The moments $\left\langle x^{m}\right \rangle$ and therefore reconstructed distribution amplitudes are valid at a given scale at which the
BSA was calculated. All our results are given for a fixed scale: $\mu = 2$~GeV.   

To conclude this section, we note again that the definition in Eq.~\eqref{LCDAdef} contains a Wilson line between the points $y_1$ and $y_2$. 
In light-cone gauge this operator is trivial: $\mathcal{W} \left[ y_2n, y_1 n \right  ] \equiv 1$, though the implementation of this gauge in numerical
approaches to bound-state equations is currently impracticable. On the other hand, it has been argued within a nonperturbative instanton vacuum 
approach~\cite{Polyakov:1997ea} that the contribution of this gauge link to the leading twist-2 quark operators is suppressed. With this in mind,
we omit the contribution of the link $\mathcal{W} \left[ y_2n, y_1 n \right  ]$ and postpone a nonperturbative approach to the matrix element.

%%%%%%%%%%%%%%%%%%%%%%%%%%%%%%%%%%%%%%%%%%%%%%%%%%%%%%%%%%%%%%%%%%%%%%%%%%%%%%%%%%%%%%%%

\subsection{Pion Distribution Amplitude}
\label{equalPDA}

In order to reconstruct $\phi_M (x , \mu)$ from the moments we write it in terms of Gegenbauer polynomials, $C_n^\alpha (2x-1)$, of order $\alpha$ which
form a complete orthonormal set on $x\in [0,1]$ with respect to the measure $[x(1-x)]^{\alpha-1/2}$. As argued in Ref.~\cite{Chang:2013pq}, the common 
projection of $\phi_M (x , \mu)$ on a $C_n^{3/2}$ $[n=0,...,\infty ]$  basis comes at the cost of a large number of terms in the Gegenbauer expansion. 
It turns out to be more economic to consider $\alpha$ itself a parameter which allows to limit the expansion to two terms for the pion, 
\begin{equation}
   \phi_\pi^\mathrm{rec.} (x,\mu) = \mathcal{N} (\alpha) \, [x \bar x]^{\alpha-1/2} \left [ 1+a_2C^\alpha_2(2x-1) \right ] \ ,
   \label{phireconst}
\end{equation}
where $\bar x = 1-x$ and the normalization is given by,
\begin{equation}
\label{pdanorm}
  \mathcal{N} ( \alpha)  = \frac{\Gamma(2\alpha+1)}{[\Gamma(\alpha+1/2)]^2}  \ ,
\end{equation}
and proceed as follows: we reconstruct the  LCDA by minimizing the function, 
\begin{align}
  \epsilon (\alpha,a_2)  = \sum^{m_\mathrm{max}}_{m=1}  &  \left | \frac{\langle x^m \rangle_\mathrm{rec.}}{\langle x^m\rangle_\pi} -1 \right|  \ , \\
  \langle x^m\rangle_\mathrm{rec.} &   = \int^1_0 dx \, x^m \phi^\mathrm{rec.}_\pi (x,\mu ) \ ,
\end{align}
with the moments $\langle x^m\rangle$ obtained by means of Eq.~\eqref{moment-BSA} and $ \langle x^m\rangle_\mathrm{rec.}$ using the
definition~\eqref{momentadef} and the expansion of Eq.~\eqref{phireconst}. We remind that in the asymptotic limit the LCDA tends to~\cite{Lepage:1979zb}: \vspace*{-2mm}
\begin{equation}
      \phi_\pi (x,\mu) \stackrel{\mu \to \infty }{=}  6x \bar x \ .
\end{equation}

%%%%%%%%%%%%%%%%%%%%%%%%%%%%%%%%%%%%%%%%%%%%%%%%%%%%%%%%%%%%%%%%%%%%%%%%%%%%%%%%%%%%%%%%

\subsection{Kaon Distribution Amplitude}

Flavored mesons like the kaon are composed of valence quarks with different masses and are not eigenstates of charge conjugation. This quark-mass
asymmetry reflects in the distribution amplitudes: $\phi_K(x) \neq \phi_K(1-x)$. In order to adapt the method described above to reconstruct the LCDA
to unequal-mass mesons, we define moments in terms of the difference of the momentum fractions denoted by,
\begin{equation}
  \xi  =x-(1-x) =  2 x-1 \ ,
\end{equation}
and define the moments,
\begin{equation}
\label{kaonmoments}
  \langle  \xi^m  \rangle_K  \  =    \int_{0}^{1} dx \, (2 x-1)^m  \phi_K (x, \mu ) \ .
\end{equation}

We thus reconstruct the kaon's LCDA with the parity decomposition,
\begin{equation}
   \phi_K^\mathrm{rec.}(x,\mu) = \phi^E_K (x,\mu)  + \phi^O_K (x,\mu) \ ,
\end{equation}
where we employ one and two Gegenbauer polynomials, respectively, in the even and odd components, 
\begin{subequations}
\begin{eqnarray} 
  \label{evenpda}  
  \phi^E_{K}(x,\mu )    & =  &  \mathcal{N}(\alpha) \, [x \bar x]^{\alpha-\frac{1}{2}}\big [1+a_2 C^\alpha_2(2x-1) \big ] ,  \\ [0.2true cm]   
  \phi^O_{K}(x,\mu )   &  =  &  \mathcal{N}(\beta) \, [x \bar x]^{\beta-\frac{1}{2}} \big [ b_1 C^\beta_1(2x-1) \nonumber \\
                                  &  +  &   b_3 C^\beta_3 (2x-1)  \big ] \ ,
 \label{oddpda}                                        
\end{eqnarray}
\end{subequations}
and $\mathcal{N}(\alpha)$ and $\mathcal{N}(\beta)$ are both as in Eq.~\eqref{pdanorm}. The even and odd components of the distribution amplitudes are 
then determined independently by separately minimizing,
\begin{align}
  \epsilon_E  (\alpha, a_2) & =  \sum_{m=2,4,..., 2m_\textrm{max}}  \left | \frac{\langle \xi^m\rangle^E_\mathrm{rec.}}{\langle \xi^m\rangle_K } -1 \right| \ , \\
  \epsilon_O (\beta,b_1,b_3) & =  \sum_{m=1,3,..., 2m_\textrm{max}-1}   \left | \frac{\langle \xi^m\rangle^O_\mathrm{rec.}}{\langle \xi^m\rangle_K } -1 \right | \ ,
\end{align}
where the reconstructed moments $\langle \xi^m\rangle^{E,O}_\mathrm{rec.}$ are obtained with the distribution amplitudes in 
Eqs.~\eqref{evenpda} and \eqref{oddpda}.

%%%%%%%%%%%%%%%%%%%%%%%%%%%%%%%%%%%%%%%%%%%%%%%%%%%%%%%%%%%%%%%%%%%%%%%%%%%%%%%%%%%%%%%%

\subsection{Heavy Mesons and Quarkonia}

The $D$ and $B$ mesons and heavy quarkonia are treated similarly, yet we employ a different functional form for $\phi_{H}^\mathrm{rec.}$ given 
by~\cite{Ding:2015rkn}, 
\begin{equation}
\label{phi-heavy}
   \phi_H^\mathrm{rec.} (x,\mu) = \mathcal{N}(\alpha,\beta) \,  4x\bar x\,e^{4\,\alpha x\bar x+ \beta (x-\bar x)} \ ,  
\end{equation}
where the normalization is, using the definition of the error function $\operatorname{Erf}(x)  = \frac{2}{\sqrt{\pi}}\int^x_0 dt\, e^{t^2}$:
\begin{align}
   \mathcal{N}(\alpha, \beta )   = & \ 16\alpha^{5/2} \Bigg [ 4\sqrt{\alpha}\, \left ( \beta \sinh(\beta)+2a\cosh(\beta) \right )  \nonumber \\ 
           + &  \ e^{\alpha+\frac{\beta^2}{4\alpha}} \left ( -2\alpha+4\alpha^2 - \beta^2 \right ) \nonumber \\
  \times &    \left \{ \operatorname{Erf}\left(\frac{\alpha-\beta}{2\sqrt{\alpha}}\right) + \operatorname{Erf} \left( \frac{\alpha+\beta}{2\sqrt{\alpha}} \right ) \right \}  \Bigg ]^{-1} .
\end{align}
The reason for this choice is that the Gegenbauer procedure sketched above is appropriate for broader and concave amplitudes, whereas a 
distribution amplitude with a convex-concave behavior of functions reminiscent of the $\delta$-function in the infinite-mass limit is more appropriately
described by Eq.~\eqref{phi-heavy}. This functional form of the distribution amplitude for heavy quarkonia is also found in the application of 
the maximum entropy method to extract the Nakanishi weight function of the quarkonia's Bethe-Salpeter wave function~\cite{Gao:2016jka}. 

We verify the validity of our reconstruction with the simple polynomial ansatz, $\phi_H^\mathrm{rec.}  (x,\mu) =  \mathcal{N}(\alpha, \beta )  x^\alpha (1-x)^\beta$
and observe that over the entire range, $x\in [0,1]$, the LCDAs reconstructed either way are but indistinguishable. The uncertainty in reconstructing the LCDA is 
therefore much smaller than that due to the model parameter $\omega_f$. On the other hand, using the separation in even and odd components with Eqs.~\eqref{evenpda} 
and \eqref{oddpda} in case of the $D$ and $B$ mesons requires the computation of a large number of Mellin moments to fix their coefficients. The larger moments 
suffer numerical instabilities for these heavy-light mesons and we thus prefer the representation in Eq.~\eqref{phi-heavy}.

We reconstruct the LCDA as in Section~\ref{equalPDA} by minimizing, 
\begin{equation}
   \epsilon  (\alpha, \beta) = \sum^{m_\textrm{max}}_{m=1}  \left | \frac{\langle x^m \rangle_\textrm{rec.}}{\langle x^m\rangle_H } -1 \right | \ ,
\end{equation}
with $\langle x^m \rangle _\textrm{rec.}$ calculated as described before and making use of Eq.~\eqref{phi-heavy}.

%%%%%%%%%%%%%%%%%%%%%%%%%%%%%%%%%%%%%%%%%%%%%%%%%%%%%%%%%%%%%%%%%%%%%%%%%%%%%%%%%%%%

\subsection{Mellin Moments  \label{Mellin}}

The Mellin moments $\langle x^m\rangle$ are integrals over a BSA and quark propagators. We follow Ref.~\cite{Chang:2013pq} in using 
Nakanishi-type representations of the scalar BSA amplitudes $\mathcal{ F}_i= E_M$, $ F_M$, $G_M$, $H_M $, and likewise use the \texttt{2ccp} propagators~\eqref{ccp} 
which allows us to represent the moments in Eq.~\eqref{moment-BSA}  by Feynman integrals.  The amplitudes $\mathcal{ F}_i$ for equal-valence quark mesons 
are therefore parametrized by, 
\begin{equation}
\label{pionnakanishi}
  \mathcal{ F}_i (k,P)  =  \mathcal{ F}^{\rm ir}_M (k,P) + \mathcal{ F}^{\rm uv}_M(k,P),
\end{equation}
with the definitions,
\begin{eqnarray}
\label{calFirfit}
\mathcal{ F}^{\rm ir}_M(k,P) & = & c_\mathcal{ F}^{\rm ir}\int_{-1}^1 \! dz \, \rho_{\nu^{\rm ir}_\mathcal{ F}}(z) \bigg[
                                                       a_\mathcal{ F} \hat\Delta_{\Lambda^{\rm ir}_{\mathcal{ F}}}^4(k_z^2)  \nonumber  \\
                                             &  & \hspace*{3mm}+ \, a^-_\mathcal{ F} \hat\Delta_{\Lambda^{\rm ir}_\mathcal{ F}}^5(k_z^2) \bigg ] \ ,
\end{eqnarray}
\begin{eqnarray}
\label{Euvfit}
E^{\rm uv}_M(k,P) & = & c_{E}^{\rm uv} \int_{-1}^1 \! dz \, \rho_{\nu^{\rm uv}_E}(z)\,   \hat \Delta_{\Lambda^{\rm uv}_{E}}(k_z^2)  \, ,  \\
\label{Fuvfit}
F^{\rm uv}_M(k,P) & = & c_{F}^{\rm uv} \int_{-1}^1 \! dz \, \rho_{\nu^{\rm uv}_F}(z)\, \Lambda_F^{\rm uv} k^2 \Delta_{\Lambda^{\rm uv}_{F}}^{2}(k_z^2)\, ,   \\
\label{Guvfit}
G^{\rm uv}_M(k,P) & = & c_{G}^{\rm uv} \int_{-1}^1 \! dz \, \rho_{\nu^{\rm uv}_G}(z)\,
 \Lambda_G^{\rm uv}\Delta_{\Lambda^{\rm uv}_{G}}^{2}(k_z^2)  \, , 
\end{eqnarray}
where $\hat \Delta_\Lambda(s) = \Lambda^2 \Delta_\Lambda(s)$, $\Delta_\Lambda(s) = 1/(k^2_z+\Lambda^2)$, $k_z^2=k^2+ z k\cdot P $, $a^-_E = 1 - a_E$, 
$a^-_F = 1/\Lambda_F^{\rm ir} - a_F$,  $a^-_G = 1/[\Lambda_G^{\rm ir}]^3 - a_G$, and the spectral density is given by,
\begin{equation}
    \rho_\nu = \frac{\Gamma\left(\nu+3/2\right)}{\sqrt{\pi}\Gamma\left(\nu+1\right)}(1-z^2)^\nu.
 \label{rhropion}   
\end{equation}
The scalar amplitude $H(k,P)$ is negligibly small, has little impact, and is thus neglected. We do not fit the amplitudes directly but rather the Chebyshev moments
$\mathcal{F}_i^m ( k, P )$ of the expansion,  \vspace*{-2mm}
\begin{equation}
   \mathcal{F}_i ( k, P ) = \sum_{m=0}^{\infty} \mathcal{F}_i^m ( k, P ) U_m (z_p ) \ ,
\end{equation}
where $z_p = k\cdot P/|k||P|$ and we typically use $m=4$ Chebyshev polynomials $U_m (z_p )$. 

The fit parameters for mesons with equal valence-quark masses, namely $\pi, \eta_c$ and $\eta_b$, are tabulated in Tab.~\ref{pi-Nakanishi}, Appendix~\ref{App-A}.
We here report the  first three Mellin moments computed  with Eq.~\eqref{moment-BSA} combining the \texttt{2ccp} parametrization for the quark propagators in 
Tab.~\ref{tab:2ccp}  and the Nakanishi representation  of the BSAs introduced above.\medskip

\begin{center}
\begin{tabular}{c|c|c|c}
\hline \hline
$\langle x^m\rangle_M$&$\langle x\rangle$&$\langle x^2\rangle$&$\langle x^3\rangle$\\ [0.7mm]
\hline
$\langle x^m\rangle_\pi$   &  0.500& 0.318$\pm$0.008& 0.228$\pm$0.006\\
$\langle x^m\rangle_{\eta_c}$  & 0.500& 0.273$\pm$0.001& 0.160$\pm$0.001\\
$\langle x^m\rangle_{\eta_b}$  & 0.500&0.262$\pm$0.001& 0.144$\pm$0.001\\
\hline \hline
\end{tabular}
\end{center}\medskip

In the case of flavored mesons with unequal quark masses, a satisfactory representation of the numerical solutions to the scalar function of the BSAs 
is~\cite{Shi:2014uwa},
\begin{equation}
 \label{BSAparKD}
 \hspace*{-2mm}
   \mathcal{F}_i (k,P)=\sum^{2}_{\sigma=0}  \,  \int^1_{-1}  \!dz\,  \rho_{\nu_\sigma}(z) 
       \frac{\mathcal{U}_\sigma\Lambda^{2n_\sigma}_{\mathcal{F}_i}}{(k^2+z\, k\cdot P + \Lambda^2_{\mathcal{F}_i})^{n_\sigma}} \, ,
\end{equation}
using $\mathcal{U}_0 = U_0-U_1-U_2$, $\mathcal{U}_1 = U_1$ and $\mathcal{U}_2 = U_2$. The set of parameters that fit the kaon BSA
is listed in Tab.~\ref{K-Nakanishi} of Appendix~\ref{App-A}. We calculated the Mellin moments of the kaon $\langle \xi^m\rangle_K$  using Eq.~\eqref{moment-BSA},
where we remind that the moments are in terms of $\xi = 2x-1$, as we separated the even and odd moments to reconstruct the LCDA. The first three 
moments are:
\medskip

\begin{center}
\begin{tabular}{c|c|c|c}
\hline \hline
 $\langle \xi^m \rangle$    &  $\langle \xi \rangle$  &  $\langle \xi^2\rangle$ &  $\langle \xi^3\rangle$\;    \\ [0.7mm]
\hline
$\langle \xi^m\rangle_K$    &  0.124$\pm$0.013  & 0.234$\pm$0.006  & 0.068$\pm$0.005\\
\hline \hline
\end{tabular}
\end{center}  \medskip

On the other hand, we compute the  moments $\langle x^m\rangle_M$ of the $D$ and $B$ mesons since we do not separate the even
and odd components of the distribution amplitude by means of Gegenbauer polynomials. In principle, this can also be achieved, yet 
while we manage to obtain a reasonable fit for the $D$ and $D_s$, the solutions for the $B$ mesons are numerically unstable. 
This is not unexpected, as the distribution amplitudes of the heavy-flavored mesons reveal a pronounced asymmetry with respect to $x \to (1-x)$
not easily reproduced with just a few Gegenbauer polynomials.  The BSA of the $D$ and $B$ mesons are fitted to the Nakanishi-like representation 
in Eq.~\eqref{BSAparKD} and also tabulated in Tabs.~\ref{DuNakanishi} to \ref{BcNakanishi} of Appendix~\ref{App-A}. 
\medskip

\begin{center}
\begin{tabular}{c|c|c|c}
\hline \hline
$\langle x^m\rangle_M$   &  $\langle x\rangle$&$\langle x^2\rangle$&$\langle x^3\rangle$\;  \\ [0.7mm]
\hline
$\langle x^m\rangle_{D_u}$&0.633$\pm$0.007& 0.447$\pm$0.009& 0.334$\pm$0.011\\
$\langle x^m\rangle_{D_s}$&0.578$\pm$0.009& 0.375$\pm$0.011& 0.260$\pm$0.011 \\
$\langle x^m\rangle_{B_u}$&0.833$\pm$0.005&0.707$\pm$0.008& 0.608$\pm$0.009 \\
$\langle x^m\rangle_{B_s}$&0.821$\pm$0.003& 0.689$\pm$0.006& 0.584$\pm$0.007\\
$\langle x^m\rangle_{B_c}$&0.732$\pm$0.002& 0.545$\pm$0.003& 0.414$\pm$0.003\\
\hline \hline
\end{tabular}
\end{center}\medskip

\noindent
To reconstruct the heavy-meson LCDA only these three moments are needed.

%%%%%%%%%%%%%%%%%%%%%%%%%%%%%%%%%%%%%%%%%%%%%%%%%%%%%%%%%%%%%%%%%%%%%%%%%%%%%%%%%%%%

\begin{figure*}[t!]
\centering
  \includegraphics[scale=0.85,angle=0]{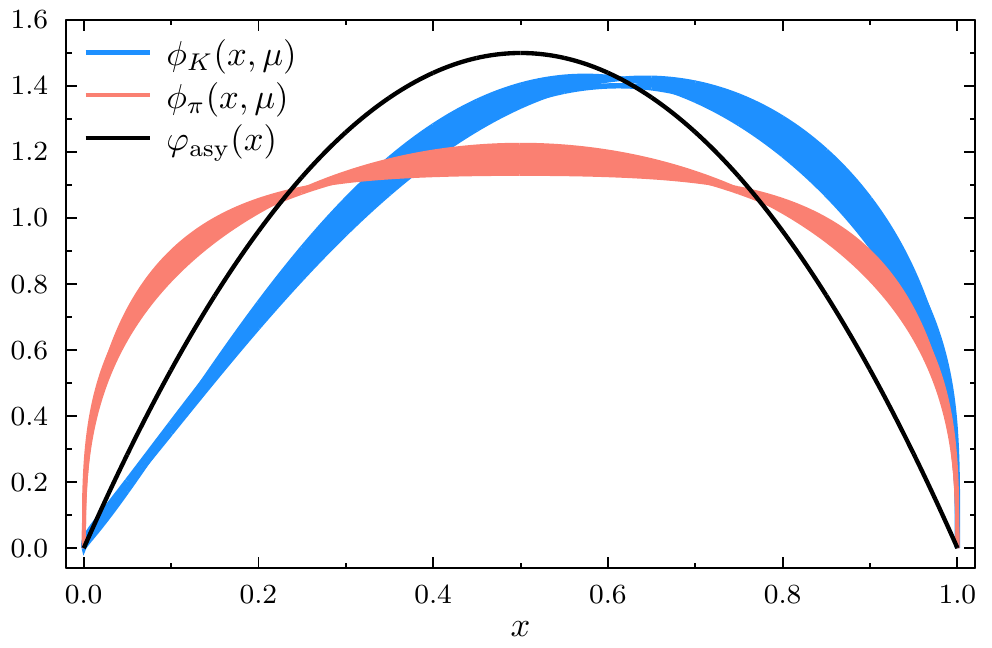}  \hspace*{6mm} 
   \includegraphics[scale=0.85,angle=0]{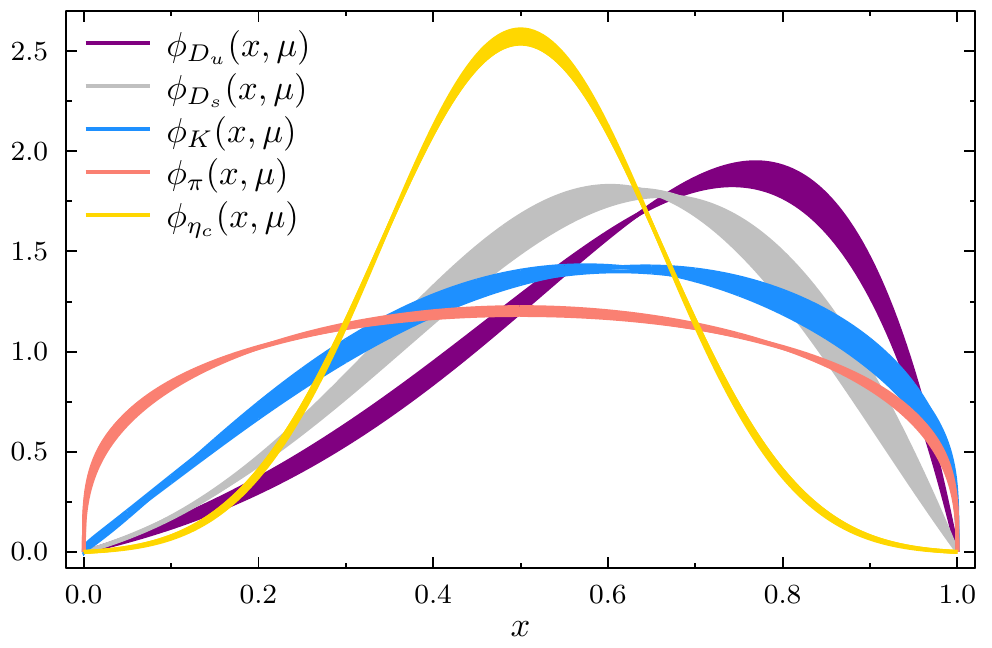}      
\caption{\label{piKDetacPDA} Distribution amplitudes on the light front at a renormalization point $\mu = 2$~GeV.  
   \textbf{Left panel}: $\phi_\pi (x, \mu )$, $\phi_K(x,\mu )$ and $\phi_\mathrm{asy} (x)=6x\bar x$ is the asymptotic LCDA. 
   \textbf{Right panel}:  Comparison of the light-meson distribution amplitudes with $\phi_{D_u} (x,\mu)$,  $\phi_{D_s} (x,\mu)$ and  $\phi_{\eta_c} (x,\mu)$.
   The error bands correspond to uncertainties of $\omega_f  \pm  \Delta\omega_f$ in the interaction model.  }
\end{figure*}

%%%%%%%%%%%%%%%%%%%%%%%%%%%%%%%%%%%%%%%%%%%%%%%%%%%%%%%%%%%%%%%%%%%%%%%%%%%%%%%%%%
%%%%%%%%%%%%%%%%%%%%%%%%%%%%%%%%%%%%%%%%%%%%%%%%%%%%%%%%%%%%%%%%%%%%%%%%%%%%%%%%%%

\section{ Reconstructed Distribution Amplitudes  \label{results}} 

We are now able to present numerical results for the reconstructed LCDAs of the pseudoscalar mesons discussed in Section~\ref{pda}.  The economic form of 
Eq.~\eqref{phireconst} limited to two terms in the Gegenbauer expansion  can  be fitted with  $m_{\rm max}=50$ moments, $\langle x^m\rangle_\pi$, and  yields 
the parameters: \medskip

\begin{center}
\begin{tabular}{c|c|c}
  \hline\hline
& $\alpha$ & $a_2$  \\ [0.5mm]
\hline
 $\pi$&0.867$\pm$0.023& $-0.022 \pm$0.030\\  [0.5mm]
    \hline\hline
\end{tabular}
\end{center} \medskip

\noindent
The errors are due to the theoretical uncertainty $\Delta \omega_u = \pm 0.05$ of the interaction model in setting the scale with the pion and kaon mass. 

Likewise we obtain the parameters for the kaon LCDA with Eqs.~\eqref{evenpda}, \eqref{oddpda},  and $m_{\rm max}=60$ moments  $\langle x^m\rangle_K$,
where 30 moments are even and 30 are odd:
\medskip

%%%%%%%%%%%%%%%%%%%%%%%%%%%%%%%%%%%%%%
\begin{center}
\begin{tabular}{c|c|c}
    \hline\hline 
  $K$  &  $\alpha$  & $a_2$  \\ [0.5mm]
\hline
 Even &0.839$\pm$0.049& $-0.174\pm$0.036 \\ [0.5mm]
   \hline  \hline
\end{tabular}
%%%%%%%%%%%%%%%%%%%%%%%%%%%%%%%%%%%%%%
\smallskip  \\
%%%%%%%%%%%%%%%%%%%%%%%%%%%%%%%%%%%%%%
\begin{tabular}{c|c|c|c}
    \hline\hline
    $K$  &$\beta$&  $b_1$  &   $ b_3$  \\     [0.5mm]
\hline            
Odd& 0.817$\pm$0.041&0.277$\pm$0.031&0.015$\pm$0.009 \\  [0.5mm]
    \hline\hline
\end{tabular} 
\end{center}
%%%%%%%%%%%%%%%%%%%%%%%%%%%%%%%%%%%%%%
\medskip

The heavy quarkonia and heavy-light mesons, for which we use an exponential parametrization for the distribution amplitude~\eqref{phi-heavy}
and  $m_{\rm max}=3$ moments, are described with the following parameter set:  \medskip

%%%%%%%%%%%%%%%%%%%%%%%%%%%%%%%%%%%%%%
\begin{center}
\begin{tabular}{ C{1cm}| C{2cm}| C{2cm} }
    \hline\hline
    &  $\alpha$  &  $\beta$         \\ [0.5mm]
\hline
$D_u$ &  0.038$\pm$0.005 &1.431$\pm$0.085   \\  
$D_s$ &  0.712$\pm$0.157 &  0.929$\pm$0.082 \\
$\eta_c$ &3.940$\pm$0.134&  0.0  \\ 
\hline
\end{tabular}
%%%%%%%%%%%%%%%%%%%%%%%%%%%%%%%%%%%%%%
\smallskip \\
\begin{tabular}{ C{1cm}| C{2cm}| C{2cm} }
\hline 
$B_u$  & 0.360$\pm$0.017 & 5.706$\pm$0.225   \\
$B_s$  & 1.205$\pm$0.526 & 6.109$\pm$0.594 \\
$B_c$  & 9.063$\pm$0.021& 10.035$\pm$0.076   \\
$\eta_b$   & 8.813$\pm$0.209& 0.0  \\
    \hline\hline
\end{tabular}
\end{center}
%%%%%%%%%%%%%%%%%%%%%%%%%%%%%%%%%%%%%%
\medskip

\noindent
The theoretical uncertainties are due to $\omega_f$ variations as described in Section~\ref{numerics}.

%%%%%%%%%%%%%%%%%%%%%%%%%%%%%%%%%%%%%%%%%%%%%%%%%%%%%%%%%%%%%%%%%%%%%%%%%%%%%%%%%%%%
\begin{figure*}[t!]
\centering
   \includegraphics[scale=1,angle=0]{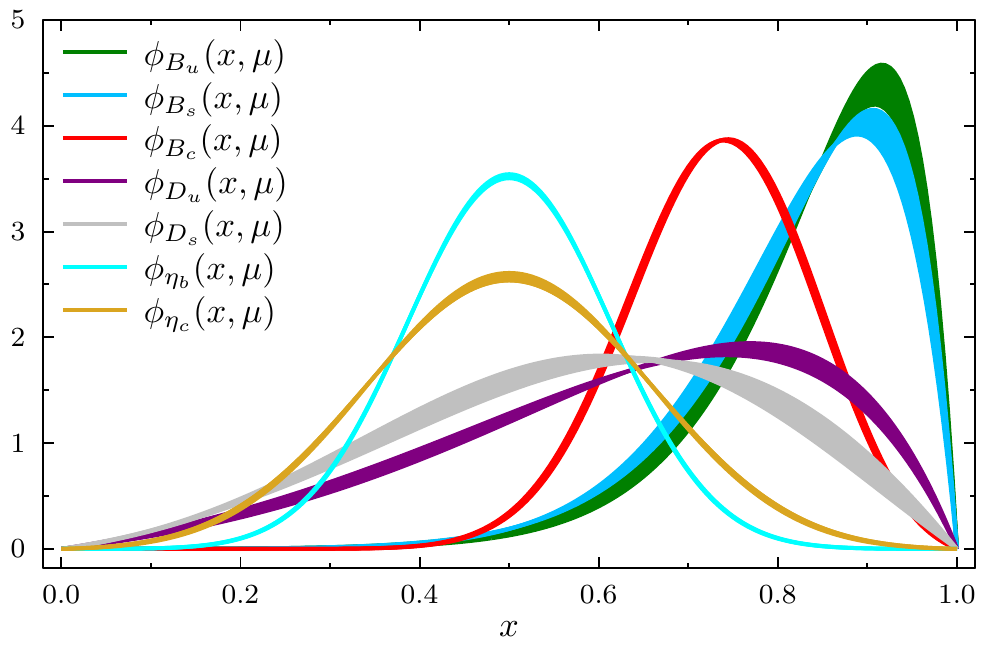}     
   \caption{\label{Heavy-PDA}  Distribution amplitudes on the light front of $D$ and $B$ mesons and the $\eta_c$ and $\eta_b$ quarkonia at a 
    renormalization point $\mu = 2$~GeV.  The error bands correspond to a variation of $\omega_f  \pm  \Delta\omega_f$ in the interaction model.   } 
\end{figure*} 
%%%%%%%%%%%%%%%%%%%%%%%%%%%%%%%%%%%%%%%%%%%%%%%%%%%%%%%%%%%%%%%%%%%%%%%%%%%%%%%

In the left panel of Fig.~\ref{piKDetacPDA} we observe that $\phi_\pi(x,\mu)$, is \emph{concave\/}, symmetric and much broader than the asymptotic limit 
$\varphi_\mathrm{asy} (x)$ as a consequence of DCSB. The symmetric shape of the pion's LCDA is precisely due to the fact that this meson is made up of 
two quarks of the same flavor, each carrying the same amount of momentum fraction of the bound state on the light front. On the other hand, $\phi_K(x,\mu)$ 
turns out to be equally concave, yet its functional form is characterized by an asymmetric shift toward a peak at $x=0.61$. This is a clear sign of dynamical 
SU(3) flavor-symmetry breaking, where the heaviest valence quark inside the kaon carries a greater amount of the meson momentum. In this case we have 
$M^E_u/M^E_s=0.73$. 

Moving our attention to mesons with larger asymmetries, the right panel of Fig.~\ref{piKDetacPDA} shows that the LCDAs of the $D_u$ and $D_s$ are not anymore 
concave as a function of $x$, rather their functional form is \emph{convex-concave\/}. The heavier charm carries most of the fraction of the meson's momentum. 
Moreover,  $\phi_{D_u}(x,\mu)$ is slightly more asymmetric and peaks higher than $\phi_{D_s}(x,\mu)$ which is due to the fact that the mass difference between 
the strange and charm quarks is smaller, i.e.:  $M^E_u / M^E_c = 0.30$ and $M^E_s/M^E_c=0.42$.  This stands in contrast to the LCDA of the $\eta_c$ which 
is symmetric about the mid-way point $x=1/2$, though much more sharply peaked than the asymptotic limit. Its behavior as a function of $x$ can be described 
by \emph{convex-concave-convex\/}.

In Fig.~\ref{Heavy-PDA} we present a comparison of the LCDAs of the charmonium, bottonium and the different $D$ and $B$ mesons. 
We note that the $B_u$ and $B_s$ distributions are extremely asymmetric and that the heavy valence quark inside the $B_u$ and $B_s$ carries almost all 
of the meson's momentum. The maxima of $\phi_{B_u}(x,\mu)$ and $\phi_{B_s}(x,\mu)$ are located at $x=0.92$ and $x=0.90$, respectively, whereas
those of  $\phi_{D_u}(x,\mu)$ and $\phi_{D_s}(x,\mu)$ are at  $x=0.76$ and $x=0.63$. The situation of the $B_c$ is somewhere in between the lighter 
$B_u$ and $B_s$ and the quarkonia, and we observe that its LCDA is less dislocated from $x>1/2$. The maximum of $\phi_{B_c}(x,\mu)$ is attained 
at the momentum fraction $x=0.74$. Moreover, we find the ratios: $M^E_u/M^E_b=0.10$, $M^E_s/M^E_b=0.12$ and  $M^E_c/M^E_b=0.32$. 
Finally, we note that $\phi_{\eta_b}(x,\mu)$ is, as expected, narrower than $\phi_{\eta_c}(x,\mu)$.

%%%%%%%%%%%%%%%%%%%%%%%%%%%%%%%%%%%%%%%%%%%%%%%%%%%%%%%%%%%%%%%%%%%%%%%%%%%%%%%%%%%%
%%%%%%%%%%%%%%%%%%%%%%%%%%%%%%%%%%%%%%%%%%%%%%%%%%%%%%%%%%%%%%%%%%%%%%%%%%%%%%%%%%%%

\section{Matching to Heavy Quark Effective Theory \label{HQET}}

The matrix element in Eq.~\eqref{LCDAdef} implies quark fields in the full theory and therefore gives rise to distribution amplitudes in QCD 
which are not directly related to those in Heavy Quark Effective Field Theory (HQET), e.g. in QCD factorization applied to weak decays of $B$
mesons~\cite{Beneke:1999br,Beneke:2000ry,Beneke:2002jn,Bauer:2000yr,Bauer:2001yt,Bauer:2005wb,ElBennich:2006yi,ElBennich:2009da,Leitner:2010fq}. 
In HQET,  the LCDA $\phi_B(x,\mu)$ is defined by~\cite{Grozin:1996pq},
\begin{equation}
   \langle 0   | \bar{u}  ( z n ) \, \mathcal{W} \left[ z, 0 \right  ] \gamma \cdot n\, \gamma_{5}\,  h_v ( 0) | \bar B ( v) \rangle  \ ,  
\end{equation}
where $h_v$ is the heavy-quark field in the effective theory. In the heavy-quark limit, $m_Q \to \infty$, the velocity of the heavy quark is almost unaffected by the 
interactions since $\Delta v =\Delta p/m_Q$. The interaction with a light quark alters its on-shell four-momentum, $p_\mu=m_Q v_\mu$, to an off-shell momentum 
$p_\mu=m_Q v_\mu +k_\mu$, where $k \sim \Lambda_\mathrm{\tiny QCD}$ is the residual momentum. In this limit, the heavy-quark propagator reads at leading order,
\begin{equation}
  S_Q (p) =  \frac{\gamma\cdot p + m_Q}{p^2-m_Q^2}\,  \stackrel{m_Q\to \infty}{\xrightarrow{\hspace*{8mm}}}\,  \frac{1+\gamma\cdot v }{2\, v \cdot k} 
                    +  \mathcal{O}\left (\frac{k}{m_Q}\right ) ,
 \label{mqpropagator}
\end{equation}
where $v _\mu$ is a time-like unit vector, $v^2 = 1$, for instance $v _\mu=(1,\vec 0)$ in the heavy quark's rest frame. This propagator must then be inserted 
in the bound-state equation~\eqref{BSE} and the resulting BSA is  projected on the light front. 

We hold off this calculation for the time being and turn our attention instead to the inverse moment of the heavy-meson distribution amplitude, $\lambda_H(\mu)$, 
defined by,
\begin{equation}
   \frac{1}{\lambda_H (\mu)} = \frac{1}{m_H}  \int^1_0 dx \  \frac{\phi_{H} (x,\mu)}{x} \ ,
\end{equation}
which plays an important role in calculations of exclusive $B$ decays within  HQET, for example in the radiative leptonic decays 
$B\to \gamma \ell \nu_\ell$~\cite{Beneke:2018wjp}.
%
%%%%%%%%%%%%%%%%%%%%%%%%%%%%%%%%%%%%%%%%%%%%%%%%%%%%%%%%%%%%%%%%%%%%%%%%%%%%%%%%%%%
\begin{table}[t!]
\begin{center}
\begin{tabular}{ C{1cm}| C{2cm}| C{2cm} }
    \hline\hline
    & $\lambda^\textrm{\tiny QCD} (\mu)$  &  $\lambda^\textrm{\tiny HQET} (\mu)$      \\ [0.5mm]
\hline
$D_u$ &  0.391$\pm$0.010& 0.493$\pm$0.013   \\  
$D_s$ & 0.562$\pm$0.026&  0.709$\pm$0.033 \\
$B_u$ &  0.452$\pm$0.015& 0.501$\pm$0.016   \\  
$B_s$ &  0.520$\pm$0.022& 0.576$\pm$0.025   \\  
$B_c$ &  1.354$\pm$0.014& 1.501$\pm$0.016   \\  
\hline\hline
\end{tabular}
\caption{The inverse moments $\lambda^\textrm{\tiny QCD} (\mu)$ and $\lambda^\textrm{\tiny HQET} (\mu) $ in GeV at the scale $\mu=2$~GeV.}
\label{lambdaH} 
\end{center}
\end{table}
%
%%%%%%%%%%%%%%%%%%%%%%%%%%%%%%%%%%%%%%%%%%%%%%%%%%%%%%%%%%%%%%%%%%%%%%%%%%%%%%%%%%%%

We can calculate $\lambda^\textrm{\tiny QCD}_H$ using the distribution amplitude given by Eq.~\eqref{phi-heavy} in the full theory. A matching relation between 
$\lambda^\textrm{\tiny QCD}_H$  and $\lambda^\textrm{\tiny HQET}_H$  exists~\cite{Pilipp:2007sb}, which to leading order in $\alpha_s(\mu)$ is given by, 
\begin{align}
\lambda^\textrm{\tiny HQET}_H (\mu)  =  & \left  [1+\frac{\alpha_s(\mu)}{4\pi}  C_F \left (2\ln^2  \left ( \frac{\mu}{m_H}  \right )  \right. \right. \nonumber \\
             & \left. \left.+ \  4\ln \left( \frac{\mu}{m_H}\right  ) +4 + \frac{\pi^2}{12} \right ) \right ]  \lambda^\textrm{\tiny QCD}_H (\mu)  \; .
\end{align}
Here, we use for the running coupling at leading order: 
\begin{equation}
   \alpha_s(\mu)  =  \frac{4\pi}{\beta_0 \ln \left ( \frac{\mu^2}{\Lambda^2_\textrm{\tiny QCD}} \right ) } \ ,   \quad \beta_0  = 11 - \frac{2}{3}  N_f  \ .
\end{equation}
In Tab.~\ref{lambdaH} we list the inverse moments obtained with our LCDAs of the $D$ and $B$ mesons and the corresponding values in HQET at $\mu=2$~GeV.
We remind, however, that the heavy-quark expansion is not reliable in case of charmed mesons and the values for $\lambda (\mu)$ are only presented for completeness. 

For comparison, QCD sum rules predict  $\lambda_B^\textrm{\tiny HQET} (1\,\textrm{GeV})  = 0.460 \pm 0.110$\,GeV~\cite{Braun:2003wx}
and  $\lambda_B^\textrm{\tiny QCD}  = 0.460 \pm 0.160$\,GeV (no scale given)~\cite{Khodjamirian:2005ea},
a model LCDA in HQET leads to \linebreak $\lambda_B^\textrm{\tiny HQET} (2\,\textrm{GeV})  =  0.58 \pm 0.04$\,GeV~\cite{Lee:2005gza}, 
 a DSE-BSE approach finds  $\lambda_B^\textrm{\tiny QCD} (2\,\textrm{GeV})  = 0.54\pm 0.03$\,GeV~\cite{Binosi:2018rht}, whereas a range of 
$0.2\, \textrm{GeV} \leq \lambda_B^\textrm{\tiny HQET}  (1\,\textrm{GeV})  \leq 0.5\, \textrm{GeV} $ is considered in an analysis of relevant form 
factors in the decay $B\to \gamma \ell \nu_\ell$~\cite{Beneke:2018wjp}.

%%%%%%%%%%%%%%%%%%%%%%%%%%%%%%%%%%%%%%%%%%%%%%%%%%%%%%%%%%%%%%%%%%%%%%%%%%%%%%%%%%%%
%%%%%%%%%%%%%%%%%%%%%%%%%%%%%%%%%%%%%%%%%%%%%%%%%%%%%%%%%%%%%%%%%%%%%%%%%%%%%%%%%%%%

\section{Final Remarks  \label{remarks} }

Based on earlier insights in a contact-interaction model of QCD~\cite{Serna:2017nlr}, we modify the ladder truncation of the Bethe-Salpeter kernel to 
take into account the different impact of vertex dressing in case of light and heavy quarks. The usual ladder truncation works very well for heavy quarkonia,
yet earlier calculations~\cite{Rojas:2014aka} demonstrated that treating the charm and light quark on equal footing leads to issues with the hermiticity 
of the interaction kernel for heavy-light mesons. 

In essence, we keep the light-meson ladder kernel which preserves the axialvector WTI unchanged, but modify the dressing function of the charm and bottom
quark with the ansatz in Eq.~\eqref{RLkernel}. This prescription comes at the cost of introducing new interaction parameters, $\omega_c$, $\kappa_c$, $\omega_b$ 
and $\kappa_b$. Nonetheless, it is a justified price to pay not only for its phenomenological success of yielding masses and weak decay constant in very good agreement with 
experiment, but  also due to theoretical considerations. Indeed, in highly asymmetric $\bar Qq$ bound states dynamical effects cannot cancel each other to 
produce a symmetric dressing of both quark-gluon vertices in the BSE, and the interaction stren\-gth in the infrared region is strongly suppressed for the charm 
and bottom quarks. For self-consistency, we verify the values of weak decay constants of the $D$ and $B$ mesons with the GMOR relation, which is an expression 
of the WTI.

With these results, we project the Bethe-Salpeter amplitudes of the pseudoscalar mesons on the light front and compute moments of the corresponding LCDA. 
The pion and kaon are reconstructed from these moments with a Gegenbauer expansion, whereas we employ an exponential form of the LCDA for the heavy
quarkonia and heavy-light mesons. The latter assumption can be related to the Nakanishi weight function of the Bethe-Salpeter wave function by means of the
maximum entropy method, though the almost identical LCDA can be reconstructed from a simple polynomial ansatz. 

We stress that our results cannot be obtained without the modified flavor dependence in the heavy-quark sector, in particular numerical calculations of the quark
dressing function on the complex plane become feasible as singularities are avoided and our results are valid \emph{without any extrapolations\/}. The distribution 
amplitudes we compute follow the expected pattern, i.e. the pion distribution amplitude is a concave function, much broader than the asymptotic one. The same is 
observed for the kaon which in addition is not symmetric about the midpoint $x=1/2$, a visual expression of SU(3) flavor breaking due to DCSB, and this asymmetry 
is growing with increasing mass of the heavier quark. The distribution amplitudes of $D$ and $B$ mesons describe a \emph{convex-concave} function, whereas for 
the $\eta_c$ and $\eta_b$ the symmetric distribution amplitude is of \emph{convex-concave-convex} form which tends to a Dirac $\delta$ function in the infinite-mass 
limit.

Eventually, for applications in heavy-meson decays, their distribution amplitudes must be obtained from Bethe-Salpeter amplitudes 
in a heavy-quark expansion of the charm- or bottom-quark propagator including a careful nonperturbative treatment of the appropriate Wilson line. 
We have postponed this task for now, but calculated the inverse moments of the heavy-meson  distribution which can be related to those in HQET. 
Of course, the computation of the LCDA suitable to an effective theory, in particular for $B$ mesons, will be of great interest in reassessing branching 
fractions of semi-leptonic and non-leptonic decays in factorization approaches.

%%%%%%%%%%%%%%%%%%%%%%%%%%%%%%%%%%%%%%%%%%%%%%%%%%%%%%%%%%%%%%%%%%%%%%%%%%%%%%%%%%%%
%%%%%%%%%%%%%%%%%%%%%%%%%%%%%%%%%%%%%%%%%%%%%%%%%%%%%%%%%%%%%%%%%%%%%%%%%%%%%%%%%%%%

\section*{Acknowledgement}

B.E. benefitted from financial support by FAPESP, grant no.~2018/20218-4, and by CNPq, grant no.~428003/2018-4. F.E.S. is supported by a CAPES-PNPD postdoctoral  
fellowship, grant no.~88882.314890/2013-01 and R.C.S. was a CAPES Master's fellow. E.R. acknowledges support from “Vicerrectoría de Investigaciones e
Interacción Social VIIS de la Universidad de Nariño”, project numbers 1928 and 2172. J.C.-M appreciated the hospitality and support of the Laboratório de Física Teórica 
e Computacional in São Paulo and B.E. is grateful for support during his stay at the Centro de Investigación y de Estudios Avanzados of the Instituto Politécnico 
Nacional in Mexico City and at the Universidade de Nariño in San Juan de Pasto. This work is part of the INCT-FNA project Proc. No. 464898/ 2014-5.

%%%%%%%%%%%%%%%%%%%%%%%%%%%%%%%%%%%%%%%%%%%%%%%%%%%%%%%%%%%%%%%%%%%%%%%%%%%%%%%%%%%%
%%%%%%%%%%%%%%%%%%%%%%%%%%%%%%%%%%%%%%%%%%%%%%%%%%%%%%%%%%%%%%%%%%%%%%%%%%%%%%%%%%%%

\appendix
\section{Bethe-Salpeter Amplitude Parameters  \label{App-A}}

We here collect all the parameters relative to the BSAs of the pion, $\eta_c$ and $\eta_b$, and separately the parameters of all flavored mesons, namely the 
$D_u$, $D_s$, $B_u$, $B_s$ and $B_c$. The corresponding parametrizations are found in Section~\ref{Mellin}. Note that these parametrizations
correspond to fits to the unnormalized BSAs, $\tilde \Gamma^{fg}_M (k,P)$, and we relate them to the normalized BSA~\eqref{PS-BSA}  by the normalization, 
\begin{equation}
   \Gamma_M^{fg} (k,P)  = \mathcal{N}_M  \tilde \Gamma^{fg}_M (k,P) \ ,
\end{equation}
where $\mathcal{N}_M$ is obtained with Eq.~\eqref{nakanishinorm} or Eq.~\eqref{canonical}. Equivalently, we may calculate the Mellin moments with
the unnormalized scalar amplitudes $\mathcal{F}_i (k,P)$ and apply the condition~\eqref{zeromomentum}: $\langle x^0 \rangle \equiv 1$.

\begin{table*}[h]
\begin{tabular*}%{llcccccccc}
{\hsize}
{
l|@{\extracolsep{0ptplus1fil}}
c@{\extracolsep{0ptplus1fil}}
c@{\extracolsep{0ptplus1fil}}
c@{\extracolsep{0ptplus1fil}}
c@{\extracolsep{0ptplus1fil}}
c@{\extracolsep{0ptplus1fil}}
c@{\extracolsep{0ptplus1fil}}
c@{\extracolsep{0ptplus1fil}}
c@{\extracolsep{0ptplus1fil}}}\hline\hline
    & $c^{\rm ir}_\mathcal{F}$ & $c^{uv}_\mathcal{F}$ & $\phantom{-}\nu^{\rm ir}_\mathcal{F}$ & $\nu^{\rm uv}_\mathcal{F}$
    & $a_\mathcal{F}$ & $\Lambda^{\rm ir}_\mathcal{F}$ & $\Lambda^{\rm uv}_\mathcal{F}$   \\  \hline
$E_{\pi}$ & $\phantom{-}1.00\phantom{8}$ & $\phantom{-}0.03\phantom{15}$ &$-0.73$&$1.00$
    & $2.40$ &1.30& 1.00\\
$F_{\pi}$ & $\phantom{-}0.56\phantom{8}$ & $0.0041$ & $1.67$ &0
    & $2.09$ & 1.09 & 1.00\\
$G_{\pi}$ & $\,0.29$ & $0.0067$ & 1.27 & 0 & $6.60$ & 0.87& 1.00 \\\hline
$E_{\eta_c}$ & $\phantom{-}1.00\phantom{8}$ & $\phantom{-}0.42\phantom{15}$ & $2.81$ & 1.00
    & $0.53$ &2.34& 0.77 \\
$F_{\eta_c}$ & $\phantom{-}0.25\phantom{8}$ & $\phantom{-}0.03\phantom{15}$ & $8.93$ & 1.00
    & $1.03$ & 1.82&  0.73 \\
 $G_{\eta_c}$ & $\phantom{-}0.23\phantom{8}$ & $\phantom{-}0.03 \phantom{15}$ & $4.56$ & 1.00
    & 1.25& 1.42 & 0.92 \\\hline
$E_{\eta_b}$ & $\phantom{-}1.00\phantom{8}$ & $\phantom{-}0.74\phantom{15}$ & $17.41$ & 1.00
    & $-0.65$ &4.00& 1.00 \\
$F_{\eta_b}$ & $\phantom{-}0.13\phantom{8}$ & $\phantom{-}0.04\phantom{15}$ & $21.34$ & 1.00
    & $1.00$ & 2.55&0.82 \\\hline\hline
\end{tabular*}
\caption{ Parameters of the BSA representation, Eqs.~\eqref{pionnakanishi}--\eqref{rhropion}, for the $\pi$, $\eta_c$ and $\eta_b$.}
\label{pi-Nakanishi} 
\end{table*}
%
%%%%%%%%%%%%%%%%%%%%%%%%%%%%%%%%%%%%%%%%%%%%%%%%%%

\begin{table*}[h]
\begin{center}
\begin{tabular}{c|c|c|c|c|c|c|c|c|c|c}
\hline \hline
$K$& $\Lambda$&$\nu_0 $&$\nu_1$ &$\nu_2 $&$U_0 $& $U_1$& $10^3U_2$&$n_0$&$n_1$&$n_2$ \;\\ [0.5mm]
\hline
$E_0$&1.80 &-0.71 &/&1.00&1.00 & / &6.83&5&/ &1\\
$E_1$&1.95&0.24&/ &0&0.74&/ &0.36&8&/&2\\\hline
$F_0$&1.55&1.24&/&0&0.42&/&0.90&5&/&1\\
$F_1$&1.71&4.79&/&0&0.20&/ &0.01&8&/&2\\\hline
$G_0$&2.08&1.00&$-0.53$&0&0.01&0.30& $-0.01$ &10&12 &2\\
$G_1$&1.44&$-0.16$ &/ &0&0.33&/&0.70&6&/&2\\
\hline \hline
\end{tabular}
\end{center}
\caption{ Parameters of the BSA representation in Eq.~\eqref{BSAparKD} for the kaon. }
\label{K-Nakanishi} 
\end{table*}
%
%%%%%%%%%%%%%%%%%%%%%%%%%%%%%%%%%%%%%%%%%%%%%%%%%%
%
\begin{table*}[h!]
\begin{center}
\begin{tabular}{c|c|c|c|c|c|c|c|c|c|c}
\hline \hline
$D_u$& $\Lambda$&$\nu_0 $&$\nu_1$ &$\nu_2 $&$U_0 $& $U_1$& $10^3U_2$&$n_0$&$n_1$&$n_2$ \;\\ [0.5mm]
\hline
$E_0$&2.64&2.24&3.00&6.00&1.00 & $-0.27 $&60&8&4&2\\
$E_1$&2.43&6.51&2.11&8.00&0 & $-0.36$ & $-0.80$  &10&8&2 \\\hline
$F_0$&1.90&5.24&/ &3.00&0.22&/&5.00&6&/&1\\
$F_1$&2.37&3.75&/ &5.00 &$ -0.04$ &/&0.01&8&/&2 \\\hline
$G_0$&2.27&1.00&1.74&0 &0 &0.68 & $-0.01$ &10&12 &2\\
$G_1$&2.50&4.07&/ &0 &0.18&/&0.40&10&/ &2 \\
\hline \hline
\end{tabular}
\end{center}
\caption{ Parameters of the BSA representation in Eq.~\eqref{BSAparKD} for the $D_u$.}
\label{DuNakanishi}
\end{table*}
%%%%%%%%%%%%%%%%%%%%%%%%%%%%%%%%%%%%%%%%%%%%%%%%%%
%
\begin{table*}[h!]
\begin{center}
\begin{tabular}{c|c|c|c|c|c|c|c|c|c|c}
\hline \hline
$D_s$& $\Lambda$&$\nu_0 $&$\nu_1$ &$\nu_2 $&$U_0 $& $U_1$& $10^3U_2$&$n_0$&$n_1$&$n_2$ \;\\ [0.5mm]
\hline
$E_0$&2.76&1.99&1.90&0&1.00& $-0.24$ &80&8&4&1\\
$E_1$&1.79&5.28&0.14&3.00& $-0.07$ & $-0.09$ &0.20&10&8 &2 \\\hline
$F_0$&2.07&4.62&/ &3.00&0.20&/ & $-10$ &6 &/ &1\\
$F_1$&2.49&5.00&/ &5.00& $-0.03$ &/ &0.01&8 &/ &2 \\\hline
$G_0$&2.98&1.00&$-0.80$&1.00&$-0.08$ & $-0.02$ &$-1.00$&10&12&2\\
$G_1$&2.80 &3.15&/ &3.00&0.09&/&0.10&10&/ &2 \\
\hline \hline
\end{tabular}
\end{center}
\caption{\label{} Parameters of the BSA representation in Eq.~\eqref{BSAparKD} for the $D_s$.}
\end{table*}
%
%%%%%%%%%%%%%%%%%%%%%%%%%%%%%%%%%%%%%%%%%%%%%%%%%%
%
\begin{table*}[h!]
\begin{center}
\begin{tabular}{c|c|c|c|c|c|c|c|c|c|c}
\hline \hline
$B_u$& $\Lambda$&$\nu_0 $&$\nu_1$ &$\nu_2 $&$U_0 $& $U_1$& $10^3U_2$&$n_0$&$n_1$&$n_2$ \;\\ [0.5mm]
\hline
$E_0$&2.91&50.29&8.00&0&1.00 &0.38&10 & 12 & 8&2\\
$E_1$&2.45&32.60&17.86&3.00& 0.002 & $-0.33$  & 0.70 & 8& 10&2 \\\hline
$F_0$&1.89 &7.70 &12.36 &0 &0.09 &0.13 &0.05 &8 &6 &1\\
$F_1$&2.65&14.32&/&0 & $-0.01$ & / & 0.05& 8& /&2 \\\hline
$G_0$&2.51&13.00 &16.82 & / & $-0.06$ &0.90 &/ &10& 12&/\\
$G_1$&2.85&23.10&/ &3.00&0.06 &/ & 0.1& 10& /&2 \\
\hline \hline
\end{tabular}
\end{center}
\caption{\label{} Parameters of the BSA representation in Eq.~\eqref{BSAparKD} for the $B_u$.}
\end{table*}
%
%%%%%%%%%%%%%%%%%%%%%%%%%%%%%%%%%%%%%%%%%%%%%%%%%%
%
\begin{table*}[h!]
\begin{center}
\begin{tabular}{c|c|c|c|c|c|c|c|c|c|c}
\hline \hline
 $B_s$& $\Lambda$&$\nu_0 $&$\nu_1$ &$\nu_2 $&$U_0 $& $U_1$& $10^3U_2$&$n_0$&$n_1$&$n_2$ \;\\ [0.5mm]
\hline
$E_0$&2.14 & 15.43&11.00 &0&1.00&1.36& 9.00 &10&6 &1\\
$E_1$&2.55&29.34&16.34 &3.00 & $-0.16$ &  $-0.40$ &0.70& 10& 8& 2\\\hline
$F_0$&1.87 &11.00 &10.42 & 0 & 0.09& 0.17 &0.05&10&6&1\\
$F_1$& 2.50&16.46&/ &0 & $-0.02$ &/  &0.05& 8&/ & 2\\\hline
$G_0$&2.64 &10.00 &18.25&0&-0.15 & 0.50 & $-1.00$ &10&12 &2\\
$G_1$& $-3.20$ &10.13 &/&3.00 &0.06& / &0.01&10& /& 2\\
\hline \hline
\end{tabular}
\end{center}
\caption{ Parameters of the BSA representation in Eq.~\eqref{BSAparKD} for the $B_s$.} 
\end{table*}
\begin{table*}[h!]
\begin{center}
\begin{tabular}{c|c|c|c|c|c|c|c|c|c|c}
\hline \hline
$B_c$ & $\Lambda$&$\nu_0 $&$\nu_1$ &$\nu_2 $&$U_0 $& $U_1$& $10^3U_2$&$n_0$&$n_1$&$n_2$ \;\\ [0.5mm]
\hline
$E_0$&2.66 &7.14&12.00 &0&1.00&0.56&5.00&6&4 &1\\
$E_1$&3.05 &35.47&17.33 &4.00& $-0.06$ & $-0.14$ &0.9 & 10&8&2 \\\hline
$F_0$&2.83 &13.53&/ &4.00 &0.08&/  &7.00&  6&/ &1\\
$F_1$& $-3.23$ &14.52&/ &5.00& $-0.01$ & / & 0.03 & 8& /& 2\\\hline
$G_0$&3.26 &8.14&14.06&/& $-0.04$ & 0.20& / &10&12 &/\\
$G_1$& $-3.56$ &13.39&/&5.00&0.03& / & 0.2& 10&/ &2 \\
\hline \hline
\end{tabular}
\end{center}
\caption{ Parameters of the BSA representation in Eq.~\eqref{BSAparKD} for the $B_c$.} 
\label{BcNakanishi}
\end{table*}

%%%%%%%%%%%%%%%%%%%%%%%%%%%%%%%%%%%%%%%%%%%%%%%%%%%%%%%%%%%%%%%%%%%%%%%%%%%%%%%%%%%%
%%%%%%%%%%%%%%%%%%%%%%%%%%%%%%%%%%%%%%%%%%%%%%%%%%%%%%%%%%%%%%%%%%%%%%%%%%%%%%%%%%%%

\end{document}